\documentclass{article}

\usepackage{arxiv}

\usepackage[utf8]{inputenc} 
\usepackage[T1]{fontenc}    
\usepackage{hyperref}       
\usepackage{url}            
\usepackage{booktabs}       
\usepackage{amsfonts}       
\usepackage{nicefrac}       
\usepackage{microtype}      
\usepackage{lipsum}
\usepackage{graphicx}
\graphicspath{ {./images/} }
\usepackage{multirow}%
\usepackage{amsmath,amssymb,amsfonts}%
\usepackage{amsthm}%
\usepackage{mathrsfs}%
\usepackage[title]{appendix}%
\usepackage{textcomp}%
\usepackage{manyfoot}%
\usepackage{booktabs}%
\usepackage{algorithm}%
\usepackage{algorithmicx}%
\usepackage{algpseudocode}%
\usepackage{listings}%
\usepackage[dvipsnames]{xcolor}%
\usepackage{makecell}%

\title{Acceleration of crystal structure relaxation with Deep Reinforcement Learning}

\author{
 Elena Trukhan \\
  Skolkovo Institute of Science and Technology\\
  Moscow, 121205 \\
  \texttt{Elena.Trukhan@skoltech.ru} \\
   \And
 Efim Mazhnik \\
  Skolkovo Institute of Science and Technology\\
  Moscow, 121205 \\
  \texttt{Efim.Mazhnik@skoltech.ru} \\
  \And
 Artem R. Oganov \\
  Skolkovo Institute of Science and Technology\\
  Moscow, 121205 \\
  \texttt{A.Oganov@skoltech.ru} \\
}

\begin{document}
\maketitle
\begin{abstract}
We introduce a Deep Reinforcement Learning (DRL) model for the structure relaxation of crystal materials and compare different types of neural network architectures and reinforcement learning algorithms for this purpose. Experiments are conducted on Al-Fe structures, with potential energy surfaces generated using EAM potentials. We examine the influence of parameter settings on model performance and benchmark the best-performing models against classical optimization algorithms. Additionally, the model's capacity to generalize learned interaction patterns from smaller atomic systems to more complex systems is assessed. The results demonstrate the potential of DRL models to enhance the efficiency of structure relaxation compared to classical optimizers.
\end{abstract}


\section{Introduction}\label{sec1}

Structure relaxation is the optimization of the positions of atoms and the lattice geometry in the process of searching for the minimum of the potential energy and interatomic forces, which corresponds to a (meta)stable state. It is a fundamental part of atomistic and molecular modeling and is used in a wide range of scientific and engineering tasks to study the properties and behavior of various materials and systems.

Many structural relaxation algorithms formalize the problem as a function minimization task \cite{VASP_Ion_opt} and search for a (meta)stable state iteratively: at each step, the potential energy, its gradients (forces), and in some methods, the Hessian matrix (stresses), are evaluated using various techniques, including first-principles methods like the density functional theory (DFT) and the Many-Body Perturbation Theory (MBPT), as well as empirical potentials, machine learning potentials, and others. The data obtained from these calculations are then used in optimization algorithms such as steepest descent \cite{steep_des}, conjugate gradients \cite{conj_grad}, Monte Carlo \cite{MonteCarlo}\cite{MonteCarlo2}, and molecular dynamics \cite{MolecularDyn}. These algorithms guide the shift of atoms toward the direction of minimum energy.

The issue arises when the complexity of the structure increases, such as with the growth in the number of atoms $N$ in a unit cell or the diversity of chemical elements. This leads to an increase in the time required for relaxation, because the computational complexity of calculating the energy and forces increases with $N$, starting from $O(N^{3})$ for DFT\cite{PhysRevB.99.064114} up to $O(N^{7})$\cite{GW_scale}\cite{HF_scale} and $\exp(2N)$ with more accurate methods. The complexity of the potential energy surface (PES) also increases with system size, resulting in a greater number of local minima that the relaxation process must navigate \cite{FHI_relax}.

Deep Reinforcement Learning (DRL) can be used in this case to predict the most efficient steps of optimization and to overcome the mentioned time problem by reducing the number of steps. In the field of material science, Reinforcement Learning has been used mainly for the design of new materials \cite{pan2022deep}\cite{MatDes}\cite{article_clast}\cite{Zhou2019-uz}, synthesis planning \cite{Wang2020} and microstructure optimization \cite{Vasudevan_2022} but relaxation based on RL is promising for several reasons:
\begin{itemize}
\item It enables the utilization of experience accumulated over a large number of optimizations, which are ignored by existing optimization methods. This is particularly beneficial because many tasks involve the relaxation of structures that have already been optimized, allowing us to leverage known paths from disturbed to relaxed structures. For example, studying the catalytic properties of a material involves the relaxation of structures that differ only in the position of several molecules on the surface.
\item RL addresses the challenge of generating representative datasets, as agent gathers data directly by interacting with the environment. 

\item The objective of reducing the number of relaxation steps can be established both directly by selecting the reward function as a function of the number of steps and indirectly through the introduction of a discount parameter. 
\end{itemize}

In the presented work, a Deep Reinforcement Learning (DRL) model for the structure relaxation process is introduced. Crystal structures are represented as crystal graphs, following the methodology of Xie et al. (2018) \cite{Xie_2018}.  Two types of neural networks are tested in this work: Crystal Graph Convolutional Neural Networks (CGCNN) \cite{Xie_2018} and E(3)-Equivariant Tensor Field Networks \cite{E3}. The Twin Delayed Deep Deterministic Policy Gradient (TD3) \cite{fujimoto2018addressing} algorithm is primarily utilized for model training. However, we also tested the Soft Actor-Critic (SAC) \cite{haarnoja2018soft}, with results presented in Appendix \ref{SAC_sec}.

We compare different architectures and RL algorithms taking as a benchmark the Al-Fe system described by EAM potentials \cite{Fe_EAM}\cite{Al_EAM}\cite{AlFe_EAM} used for the PES generation.  First of all, dependence of the model performance on parameter settings is studied. Next, we compare the best models with classical optimizers: Broyden-Fletcher-Goldfarb-Shanno \cite{BFGS_art}\cite{BFGS_art2}\cite{BFGS_art3}\cite{BFGS_art4} and Conjugate gradient \cite{CG_art}. Finally, the model's ability to generalize learned patterns of interactions from systems with a small number of atoms/chemical elements to systems with a larger number of atoms/chemical elements is investigated.

\section{Methods}\label{sec2}

\subsection{Preliminary of Reinforcement Learning}\label{RL_notions}

Reinforcement learning (RL) is a field of machine learning applied to decision-making problems, where an \textit{Agent} learns to behave in a specific \textit{Environment} by performing actions and observing the outcomes. It is most commonly formulated in the form of Markov Decision Processes (MDPs). In this framework, the Agent, at a given time $t$ in state $s_t$, selects an action $a_t$. This action transitions the system to the next state $s_{t+1}$, with this transition determined by the internal dynamics of the Environment, denoted as $f(s'|s, a)$, which is also called \textit{transition law}, so  $s_{t+1} \sim f(\cdot|s_t, a_t)$. The next state $s_{t+1}$ depends solely on the current state and action, thus satisfying the Markov property. Upon transitioning, the Agent receives a reward $r_t=R(s_t,s_{t+1},a_t)$ \cite{Sutton}\cite{OpenAI}\cite{RL_ping}. 

The decision regarding which action to take is governed by the Agent's policy, denoted here as $\pi$. The main goal of RL is to find such a policy $\pi^{*}$, which is called \textit{optimal}, that maximizes an objective function $J(\pi) = \mathbb{E}_{\tau \sim \pi, f} [R(\tau)] = \mathbb{E}_{\tau \sim \pi,f} \left[ \sum_{t=0}^T \gamma^t r_t \right]$, where $\gamma \in [0,1]$ is the discount factor and $\tau = (s_0, a_0, s_1, a_1, ...)$ is the MDP trajectory,  which ends at the terminal step $T$. The initial state $s_0$ is selected randomly according to a starting distribution $\rho_0 (\cdot)$. The expected value is taken over the probability distribution $P(\tau|\pi)=\rho_0(s_0) \prod_{t=0}^{T-1}f(s_{t+1}|s_t,a_t )\pi(a_t|s_t)$. 

In the context of algorithms used in this work the Agent maximizes not the objective $J(\pi)$, but the action-value function $Q^{\pi^*}(s,a)$. In the case of Twin Delayed DDPG (TD3) the policy $\pi$ is deterministic and $Q^{\pi^*}(s,a)$ is given by the formula: 

\begin{equation}\label{BellmanOpt}
    Q^{\pi^*}(s,a) = \mathbb{E}_{s' \sim P(\tau|\pi^*)}\left[R(s, s',a) + \gamma \max_{a'} \left[Q^{\pi^*}(s',a') \right]\right]
\end{equation} 

In TD3 two classes of deep learning models are introduced. The first one, called \textit{Actor}, denoted as $\pi_{\theta}(s)$ with tunable parameters $\theta$, is used to approximate  optimal policy $\pi^{*}$. Another one, called \textit{Critic} and denoted as $Q_{\phi} (s,a)$ with tunable parameters $\phi$ approximates optimal action-value function $Q^{\pi^*}(s,a)$. One can find more details regarding corresponding loss functions in Appendix \ref{RL_ALG}. 

\subsection{Crystal structure relaxation as a Markov decision process}

The structure relaxation operates within a framework of Markov decision process MDP(\textit{S},\textit{A},\textit{R}), where we define each term in what follows: 

\begin{itemize}
    \item \textbf{\textit{S} is the state space}: each $s_t \in S$ is a crystal graph \cite{Xie_2018} of the corresponding structure at time step $t$. In this graph, each node $i$ represents an atom in the given unit cell and contains a node feature vector $v_{i}$. The $k$-th edge between two nodes $(i,j)_k$ corresponds to the bond between the $i$-th atom and the $k$-th image of its $j$-th neighbor and contains an edge feature vector $u_{(i,j)_k}$.    
    \item \textbf{\textit{A} is the action space}: each $a_t \in A$ is a graph in which each node $i$ corresponds to the node $i$ in $s_t$ and contains the vector $\Delta\vec{r}_i$, representing the atomic shift: $a_t = \{\Delta\vec{r}_1, \Delta\vec{r}_2, ..., \Delta\vec{r}_N\}$, $N$ is the number of atoms in a unit cell of the structure.  
    \item \textbf{\textit{R} is the reward function}. Four options are considered in the given work. The first one, labeled \textit{"force"}, optimizes atomic forces: 
    
    \begin{equation}\label{R1_atom}
        R_1(s_t, s_t', a_t) = - \max_{n\in [1,N]}|\vec{f}_n(s_t')|, 
    \end{equation} 
    where $\vec{f}_n$ is the force acting on the $n$-th atom. 
    
    The second option, labeled \textit{"step"}, optimizes the number of steps directly: 
    \begin{equation}\label{R2_atom}
     R_2(s_t, s_t', a_t) =
    \begin{cases}
        -1,& \text{if } d_t =\text{False } \\
        0,              & \text{otherwise}
    \end{cases}
    \end{equation}
    where $d_t$ is a \textit{"done"} flag, which is "True" when the step $t$ is terminal and "False" otherwise. 
    
    The third option, called \textit{"log force"}, is more sensitive to the change of atomic forces in close vicinity to the local minimum: 

    \begin{equation}\label{R3_atom}
        R_3(s_t, s_t', a_t) = - \log_{10}\left(\max_{n\in [1,N]}|\vec{f}_n(s_t')|\right)
    \end{equation} 

    The last one, labeled \textit{"hybrid"}, is a composition of all functions above with some weights $\{ w_1, w_2, w_3 \}$: 

    \begin{equation}\label{R4_atom}
        R_4(s_t, s_t', a_t) = \sum_{i=1}^3 w_i R_i(s_t, s_t', a_t)
    \end{equation} 

    \item In this work, the \textit{"done"} criterion for $d_t$ aligns with those commonly used in the classical algorithms:

\begin{equation}\label{done1}
        d_t =  \max_{n\in [1,N]}|\vec{f}_n(s_t')| \leq \epsilon_1, 
\end{equation} 
where $\epsilon_1$ is the force threshold, provided by the user. 
    
\end{itemize}

\subsection{Agent model implementation}

The models for Actor and Critic were implemented using two types of networks: 1) Crystal Graph Convolutional Neural Networks (CGCNN) \cite{Xie_2018} and 2) E(3)-Equivariant  Tensor-Filed Neural Networks (TFN) for point clouds \cite{E3}\cite{Batzner_2022}. Both architectures handle graph-structured data using convolutional operations to aggregate information from neighboring nodes, capturing the correlation between target values and local interactions.

The main difference between them is that in CGCNN both node and edge feature vectors are treated as arrays of numbers, without distinguishing them by geometric characteristics. Consequently, it can be translation-invariant if node and edge features do not depend on absolute atomic positions and invariant to the O(3) group only if all features are scalars; otherwise, invariance and equivariance are not guaranteed. 

In contrast, TFN is designed to maintain equivariance with respect to the whole E(3) group. All data in this model are treated as geometrical tensors, which are decomposed into irreducible representations of O(3). The convolutional layer is implemented as the direct product of irreducible representations, which guarantees equivariance of the architecture \cite{e3nn}. Additional information about the neural networks and graph construction can be found in Appendix \ref{NodeEdge}.

\subsection{Implementation of exploration}\label{Policy_and_exploration}

Exploration in RL is the strategy by which an Agent discovers new knowledge about its Environment, choosing actions that may not yield immediate rewards but improve future decision-making. The challenge is balancing exploration with exploitation, which involves making decisions based on the current policy. 

In our work, we implement exploration in TD3 using the \textit{adding noise to state} approach to preserve the symmetry of the system (a detailed discussion of the importance of this aspect in the context of RL-based structure relaxation is provided in Appendix \ref{Ap_noise_to_action}). For this purpose, vectors of forces in each node are rotated around randomly generated axes $\vec{n}$ to a random angle $\phi$, and the lengths of forces are modified to small values: $\vec{f}_i \rightarrow (1+\xi)\cdot (O(\vec{n}, \phi)\vec{f}_i)$, where $\xi \sim U(-\lambda, \lambda)$. Here $\lambda$  denotes the hyperparameter called \textit{noise level}. It is important to note that forces in each node are altered with the same rotation and lengthening/shortening. Two settings for the noise parameter $\lambda$ are utilized. In the first case $\lambda$ is constant ($\lambda = c$), while in the second case  $\lambda$ varies during the training episode according to the formula $\lambda(t) = (c_2 - c_1)(t/L) + c_1$, where $L$ is the maximum number of steps per episode ($\lambda \in [c_1, c_2]$).

During the training it was observed that the TD3 algorithm with exploration, described above, tends to get stuck in states with low forces, failing to shift structures further towards a minimum (see Fig. \ref{fig:Stuck} in Appendix \ref{Ap_greedy}, black trajectory). The Agent converges to a policy that predicts infinitesimally small actions for such states and relaxation oscillates between the same configurations. To address this issue, we proposed to include \textit{additional greedy exploration}. 

In this approach greedy actions are introduced by adding noise to a state with a very high $\lambda$, to facilitate relevant exploration for the Agent. Greedy actions are sampled if $\max_{n\in [1,N]}|\vec{f}_n(s_t')| > f_{max}$ and   $\max_{n\in [1,N]}|\vec{\Delta r}_n| < \Delta r_{max}$ for $N_{gr}$ steps during training, where $N_{gr}$, $ \Delta r_{max}$ and $f_{max}$ are tunable parameters. A more detailed comparison of Agents with and without greedy exploration is provided in Appendix \ref{Ap_greedy}.

\begin{figure} 
	\includegraphics[width=\linewidth]{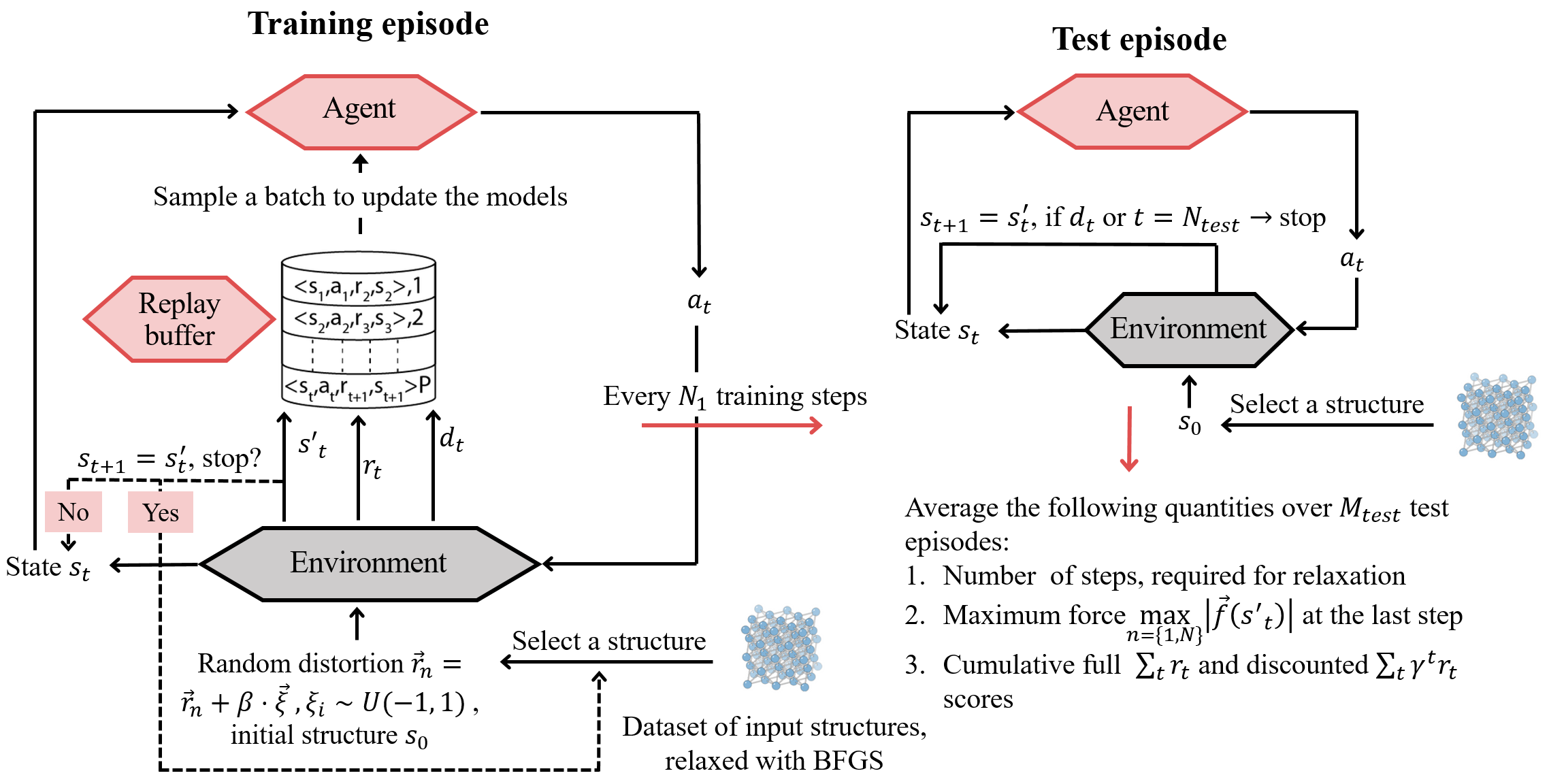}
	\caption{Workflow of the Reinforcement Learning algorithm used for structure relaxation.}
	\label{fig:Workflow}
\end{figure}

\subsection{Algorithm workflow}\label{Algorithm}

The class for the Environment is implemented according to the standards of \textit{Gym} \cite{Gym} Python library and constructed with Python Materials Genomics (\textit{pymatgen})\cite{pymatgen} and Atomic Simulation Environment (\textit{ASE})\cite{ase} libraries. 

The training loop of the RL algorithm used for structure relaxation is illustrated in Figure \ref{fig:Workflow}. Before initiating the training of the Agent, the structures in the provided dataset undergo relaxation with a classical algorithm, such as BFGS. 

The Agent undergoes training across $M_{\textrm{train}}$ episodes, each of which lasts a maximum of $N_{\textrm{train}}$ steps. At the beginning of each episode, a crystal structure is selected from the data set and disturbed from equilibrium by introducing random distortions into the atomic coordinates: $\vec{r}_n(s_0) = \vec{r}^{\;*}_n + \beta \cdot \vec{\xi}_n$, where $(\vec{\xi}_n)_i \sim U(-1,1)$, $\vec{r}_n^{\;*}$  are the coordinates of the  $n$-th atom in equilibrium. Here, $\beta$ is referred to as the \textit{distortion parameter}. This distorted state serves as the initial point in the trajectory. Then, at each step the Agent takes actions, shifts atoms in the structure, gets the next state $s_{t}'$ and \textit{"done"} flag $d_t$, stores this transition in the storage called \textit{Replay Buffer} and, if there are enough transitions in the Replay Buffer, updates Actor and Critic models. The episode ends when \textit{"done"} criterion is met or when the time limit $N_{\textrm{train}}$ is reached.

Every $N_{\textrm{1}}$ time steps training is paused to conduct $M_{\textrm{test}}$ test episodes, each lasting maximum $N_{\textrm{test}}$ steps. The testing episodes are similar to the training ones, the only difference is that the models are not updated and the transitions are not stored in the Replay Buffer. During these test episodes, various metrics such as the cumulative full and discounted scores, maximum force at the last step $\max_{n\in [1,N]}|\vec{f}_n(s_T)|$ (further referred as \textit{maximum force}) and the number of steps required for relaxation (further referred as \textit{relaxation steps}) are recorded and averaged across all test episodes. 

In this work $N_{\textrm{1}} = 1000, N_{\textrm{test}} = 100, N_{\textrm{train}} = 1000$, but these parameters can be changed by the user. 

\section{Results and Discussion}\label{sec3}

Considering that this is, to our knowledge, the first work where reinforcement learning is applied to the structure relaxation task, we find that it is important to provide a detailed description of how different model settings influence the algorithm's performance for further development of this approach. Accordingly, this section is structured as follows: Sections \ref{Comp_archs} to \ref{Noise_and_gamma} cover the investigation of the method's technical aspects, while Sections \ref{Compclass} to \ref{Gener_ability} demonstrate the algorithm's performance on practical tasks. Additionally, an analysis of the influence of the discount factor and exploration level on model performance is provided in Appendix \ref{Disc_f_and_noise}. 

The list of the structures in datasets and hyperparameters is provided in Appendix \ref{Hyperparams}. 

\subsection{Comparison of TFN and CGCNN models}\label{Comp_archs}

\begin{figure}[ht] 
	\includegraphics[width=\linewidth]{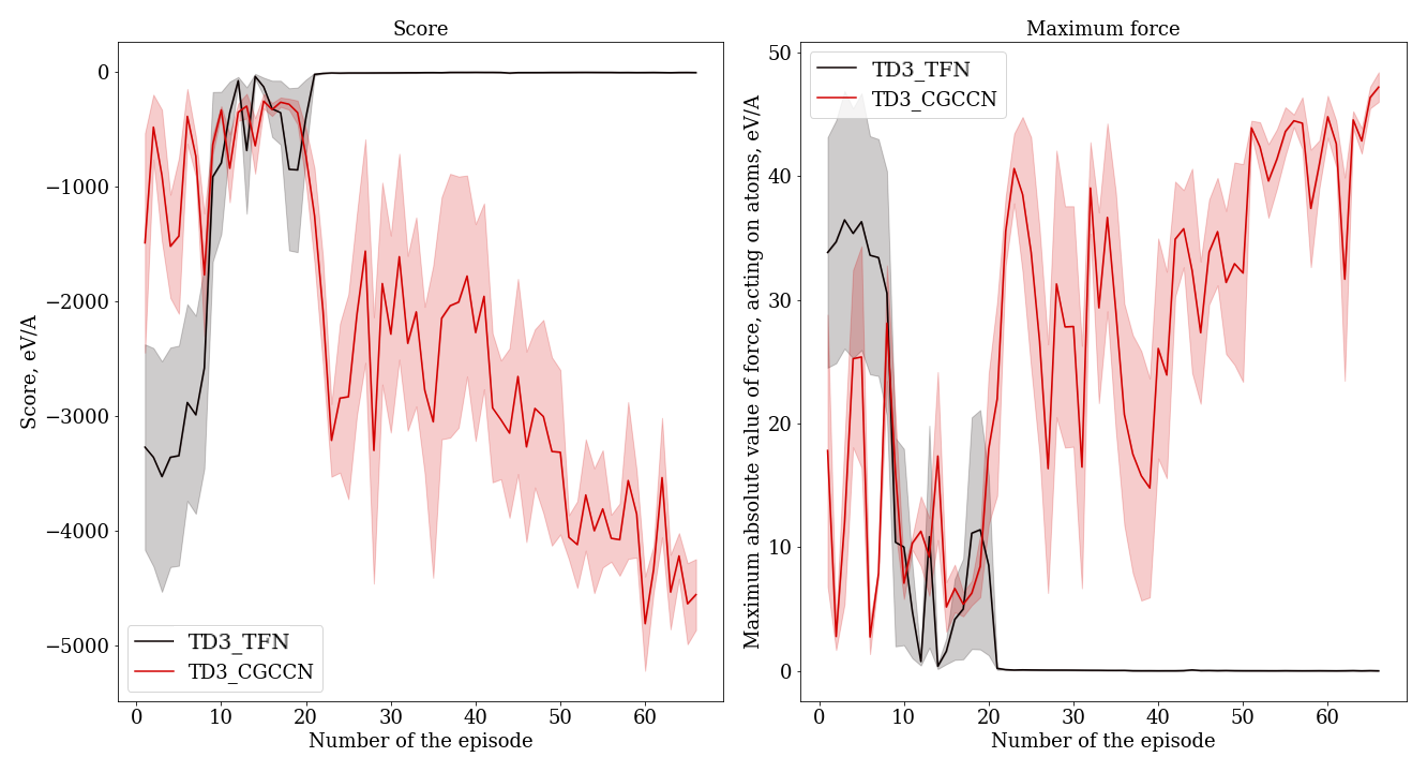}
	\caption{Learning curves of TD3 Agents with CGCNN and TFN architectures, trained on CsCl-type structure of AlFe. Curves are averaged over 5 trials with different random seeds, with the shaded area representing half a standard deviation.}
	\label{fig:CGCNN_vs_E3NN}
\end{figure}

For the comparison of two suggested architectures, the TD3 Agent was trained on CsCl-type structure of AlFe with 5 different random seeds. Averaged learning curves are presented in Fig. \ref{fig:CGCNN_vs_E3NN}. 

As can be seen, the CGCNN model is ineffective in performing the task of structure relaxation because it does not reduce the maximum force at the last step and does not increase the full score, as it should be in the case of successful training of RL Agent. This may be attributed to the continuous and high-dimensional nature of the task, both in terms of state and action spaces. From this perspective, the TFN model manages the task of relaxation better for two main reasons: firstly, it does not require training on augmented data to navigate through crystal structures differing only in rotations; secondly, the TFN model predicts actions not from the whole space of all possible atomic shifts, as in the case of CGCNN, but from a subspace of shifts restricted by the symmetry of the structure.  It has been proven that any E(3)-equivariant model obeys Curie’s principle \cite{E3NN_ord}, so the symmetry of the output can only be of equal or higher symmetry than the input. As a result, reducing the action space allows the model to quickly find the optimal policy. The drawback is that this feature limits our model because we cannot guarantee that geodesic path in the action space to the minimum always has higher symmetry than the symmetry of the structure. However, the investigation of this question is beyond the scope of this work.

As a result of the presented comparison, we further tested only TFN models.

\subsection{Comparison of reward functions and sensitivity issues}\label{Noise_and_gamma}

\begin{figure}[ht] 
	\includegraphics[width=\linewidth]{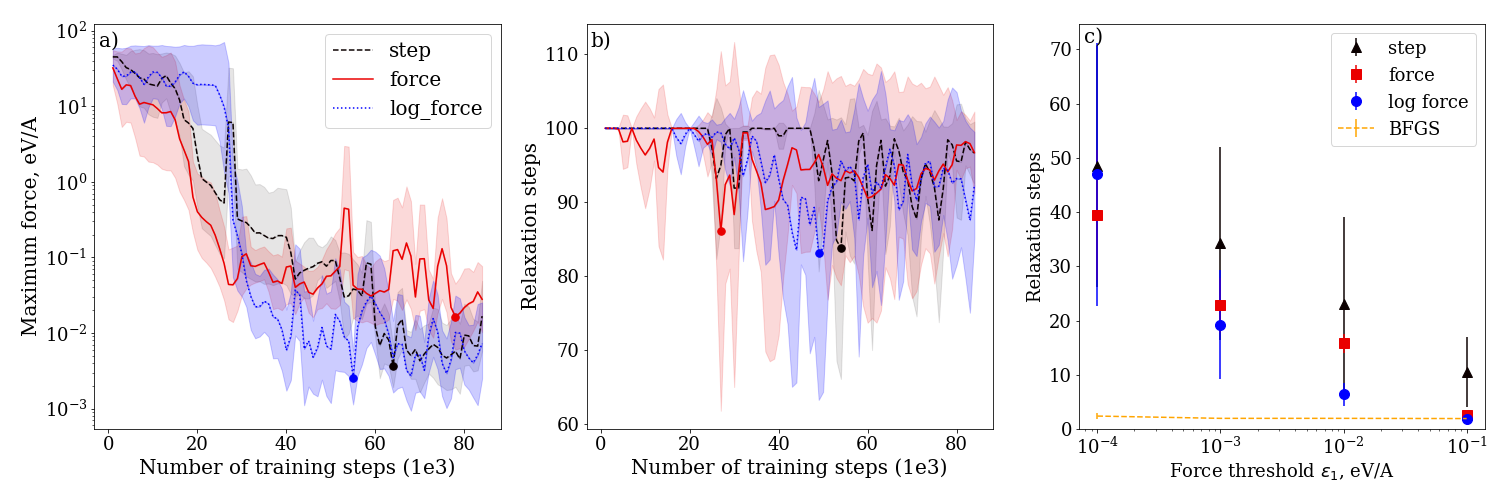}
\caption{a) - b) Learning curves of TD3 Agents with different reward functions, given by Eq. (\ref{R1_atom})-(\ref{R3_atom}). The Agents were trained on CsCl-type structure of AlFe. Curves are averaged over 5 seeds, with the shaded area representing the standard deviation. Circles mark the lowest average relaxation steps and maximum force at the last step. c) Performance of the Agents, corresponding to the models with the best results with respect to relaxation steps (corresponds to the circles on b)). The plot shows the median value with an interdecile range of a sample of 50 relaxation episodes.}
\label{fig:Different_rewards}
\end{figure}

The TD3 Agent was trained with different reward functions, outlined in Eq. (\ref{R1_atom})-(\ref{R3_atom}) on CsCl-type structure of AlFe.  The results are depicted in Fig. \ref{fig:Different_rewards}a-b). We also compare the performance of the best trained models with BFGS on relaxation up to higher values of the force threshold $\epsilon_1$  (see Fig. \ref{fig:Different_rewards}c).

First of all, one can see from Fig. \ref{fig:Different_rewards}c) that our models effectively relax CsCl-type structure of AlFe up to $\epsilon_1 = 0.1$ eV/A; however, reducing the forces further is challenging.  This issue may stem from insufficient model sensitivity near the minimum, as the differences in node and edge features of crystal structures approach the numerical error limits of TFN models and the spatial and angular resolution of the model architecture cannot be enough to distinguish them (see further discussions in Section \ref{Sens}). Consequently, adopting a \textit{log force} reward function has led to a reduction in the number of steps required for relaxation, indicating increased sensitivity for at least the reward function in the vicinity of the minimum. 

Secondly, it can be noticed that usage of the \textit{step} reward function does not provide with a better performance in terms of the number of steps required for relaxation, while from the definition of this reward function in Eq. (\ref{R2_atom}), it should optimize this value directly. This may be explained by a piecewise-constant nature of the action-value function which causes the Agent to adopt a suboptimal policy if the Agent does not face the terminal steps at the beginning of training (for the detailed discussion of this effect see Appendix \ref{step_discussion}).  

To address this issue we tested an approach where the Agent firstly is pretrained with more smooth functions, for example, \textit{hybrid} reward with $\hat{w}_1 = 1$, $\hat{w}_2 = 0$, $\hat{w}_3 = 0.5$ and then the reward function is switched to \textit{step} with consistently reducing $\epsilon_1$ in Eq. (\ref{done1}) during training. Each stage of model training continued until the Agent reached a predetermined limit on the number of steps on relaxation.

\begin{figure}[htbp]
\includegraphics[width=\linewidth]{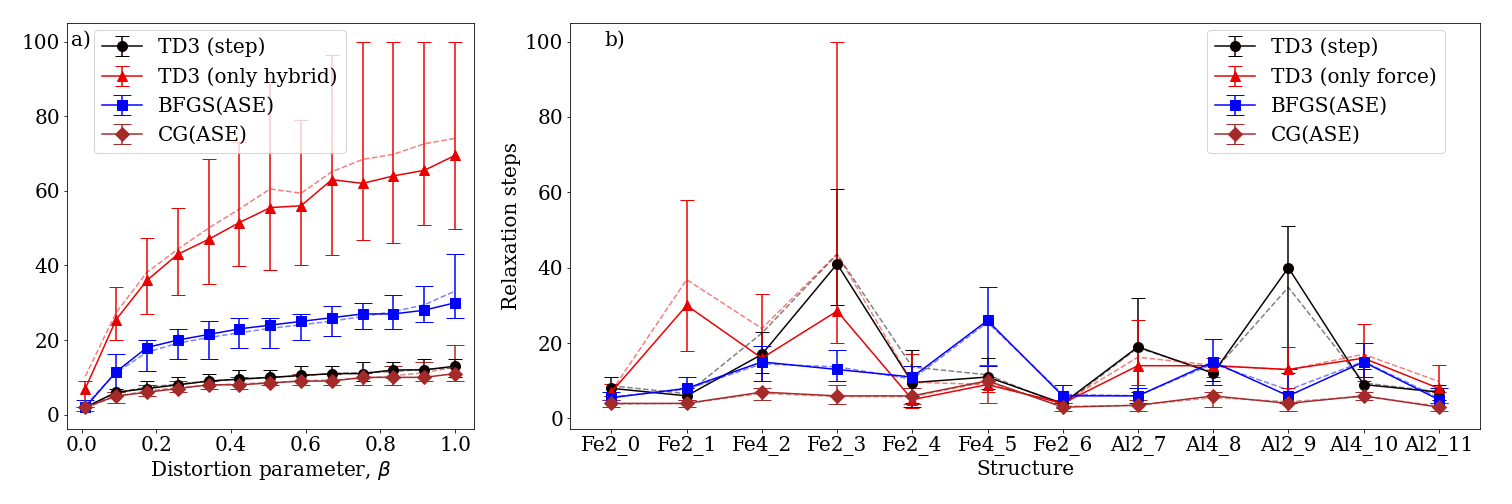}
\caption{Comparison of classical algorithms and TD3 Agents trained with and without switching to the \textit{step} reward after pertaining a) with  \textit{hybrid} reward function ($\hat{w}_1 = 1, \hat{w}_2 = 0, \hat{w}_3 = 0.5$) on hypothetical I4/mmm structure of Al b) with  \textit{force} reward function on the set of monoatomic Al and Fe structures (the description of the set is presented in Table \ref{Table_struct}). All models were trained with $\beta = 0.5$. For Al the model is tested for relaxation over a range of distortion parameter values $\beta$, displayed on the x-axis. For each point on the plots relaxation was conducted 50 times. The solid line indicates the median value with an interdecile range, while the dashed line represents the mean value.}
\label{fig:Al_step}
\end{figure}

In Fig. \ref{fig:Al_step}a-c), a comparison is presented between the performance of a model trained exclusively on the smooth reward function and one trained on the \textit{step} reward function after pretraining. This comparison is made for two cases: (1) when the Agent is trained on a single structure (the hypothetical I4/mmm structure of Al) and (2) when it is trained on a set of multiple structures simultaneously (a dataset of monoatomic Al and Fe structures). During training with the \textit{step} reward function, the parameter $\epsilon_1$ was gradually reduced— from 0.1 eV/Å to 0.01 eV/Å for the hypothetical I4/mmm structure of Al, and from 0.1 eV/Å to 0.05 eV/Å for the dataset of monoatomic Al and Fe structures.

As shown in the results, this approach enables the model to achieve performance comparable to CG when the Agent is trained on a single structure. Furthermore, Fig. \ref{fig:Al_step}a-b) demonstrates the model's ability to extrapolate its experience to higher distortions. Despite being trained with a distortion parameter $\beta = 0.5$, the model successfully relaxes structures with higher $\beta$ values during testing. However, as illustrated in Fig. \ref{fig:Al_step}c), this method does not guarantee improved results across all structures when the model is trained on a diverse dataset of different configurations. 

\subsection{Comparison with classical optimizers}\label{Compclass}

We compare the performance of the algorithms trained on structures with different numbers of atoms per unit cell $N$ using the sequential approach described in Section \ref{Noise_and_gamma}. For $N = 2$ we use results on the CsCl-type structure of AlFe from the Section \ref{Disc_f_and_noise} and for $N = 4$ — the results for hypothetical I4/mmm structure of Al from the Section \ref{Noise_and_gamma}. For $N = 8, 10, 20$ the structures are generated using USPEX code \cite{USPEX}. For each $N$ the models are trained on a single structure and tested to perform the relaxation of the same structure but randomly distorted at the beginning of the testing episode, as described in Section \ref{Algorithm}. In the case of $N = 2-8$ the relaxation during training of the models and further comparison with classical algorithms are conducted up to $\epsilon_1 = 0.01$ eV/A, while for $N = 10$ and $N = 20$ it was conducted up to $\epsilon_1 = 0.2$ eV/A and $\epsilon_1 = 0.25$ eV/A correspondingly due to the sensitivity issues mentioned in the Section \ref{Noise_and_gamma}. As one can see in Fig. \ref{fig:Gener_from_AlandFe_to_AlFe}a), the results of our algorithms are similar to classical optimizers for a small number of atoms $N$, while for higher values of $N$ it starts to outperform them.

\subsection{Generalizability of the model from simple to complex structures}\label{Gener_ability}

An important aspect of the model that we wanted to investigate was its ability to generalize interaction patterns from small systems to more complex ones. From a physical point of view, the ordering of atoms in crystal structures (except molecular crystals) is mainly conditioned by local interactions with the closest neighbors. Therefore, we can expect that an RL Agent based on GCNN will be able to extrapolate its experience from small structures to more complex ones, as the main idea of convolution is to find the correlation between target properties and local interactions. We can expect this ability to be present in two cases:

\begin{itemize}
\item The number of atoms per unit cell is increased. In this case, the ability to extrapolate means that the model effectively selects the subspace of atoms that strongly influence the position of the target node in the unit cell and accounts for the fact that, in a large structure distant atoms weakly influence each other.
\item The chemical diversity is increased. Increasing chemical diversity leads to new types of interactions that the model needs to account for, but it can generalize these patterns from its experience with other types of interactions.
\end{itemize}

\subsubsection{Chemical diversity} 

\begin{figure}[ht] 
    \includegraphics[width=1\linewidth]{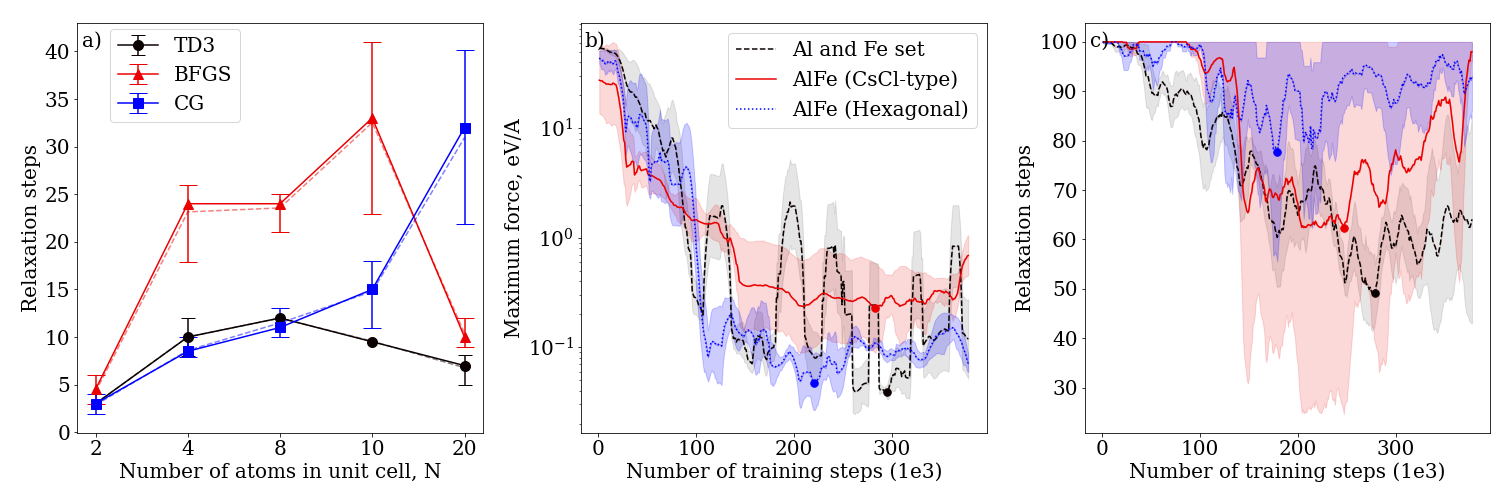}
    \caption{a) Comparison of TD3 model and classical optimizers in relaxation of structures with different number of atoms $N$ in the unit cell. For $N = 2-8$ the relaxation was conducted up to $\epsilon_1 = 0.01$ eV/A, while for $N = 10$ and $N = 20$ up to $\epsilon_1 = 0.2$ eV/A and $\epsilon_1 = 0.25$ eV/A correspondingly. For each point on the plot, relaxation was conducted 50 times. The solid line indicates the median value with an interdecile range, while the dashed line represents the mean value. b-c) Learning curve of the TD3 model, trained on the set of monoatomic Al and Fe structures with \textit{force} reward function (black) and its performance on CsCl-type (red) and simple hexagonal (blue) structures of AlFe during test. The shaded region represents a standard deviation of the average evaluation over 2 trials.} 
    \label{fig:Gener_from_AlandFe_to_AlFe}
\end{figure}

To investigate the ability of our model to extrapolate to higher chemical diversity, the model was trained on a set of monoatomic Al and Fe structures. During testing the model performed a relaxation of CsCl-type and simple hexagonal structures of AlFe without being trained on them. The results are presented in Fig. \ref{fig:Gener_from_AlandFe_to_AlFe} b-c).

As one can see, there is a strong correlation between the results for AlFe and the set of monoatomic structures, indicating that the model is able to generalize interactions between Al-Al and Fe-Fe to the case of Al-Fe interactions. The best result for the model, presented in Fig. \ref{fig:Gener_from_AlandFe_to_AlFe}c), in the case of CsCl-type structure of AlFe relaxation is $11.3 \pm 7.6$ steps, which is close to the best results of the model trained solely on this configuration (see Fig. \ref{fig:Different_gamma_and_noise_SAC_TD3}a).

\subsubsection{Number of atoms per unit cell}\label{Numbercompl}

\begin{figure}[htbp] 
    \includegraphics[width=\linewidth]{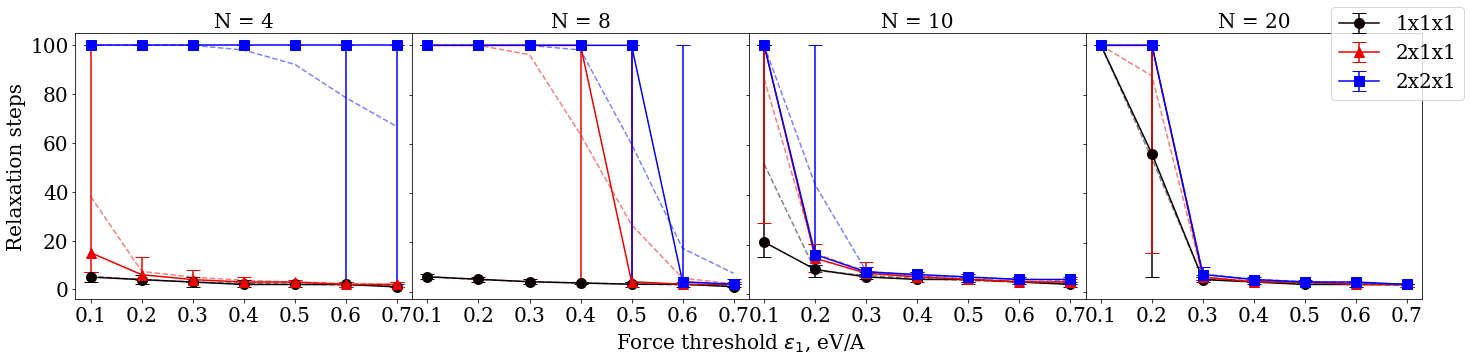}
\caption{Number of steps required for relaxation of supercells by TD3 Agents trained on structures with different number of atoms $N$ in the unit cell. Performance is measured for different $\epsilon_1$ values, displayed on the x-axis. The y-axis represents the number of steps taken by the optimizer for the relaxation of the structure, averaged over 50 runs. The solid line indicates the median value with an interdecile range, while the dashed line represents the mean value.It should be noticed that in the case of $N = 20$ the model was trained to perform the relaxation up to $\epsilon_1 = 0.25$ eV/A and for $N = 10$ up to $\epsilon_1 = 0.2$  eV/A, as it is mentioned in the Section \ref{Compclass}.}
\label{fig:Al_andFe_supercell}
\end{figure}

To investigate the ability of our model to extrapolate to a higher number of atoms per unit cell, we test how the models trained on single structures with different number of atoms per unit cell relax the corresponding supercells $2\times1\times1$ and $2\times2\times1$ without being trained on them. The results are presented in Fig. \ref{fig:Al_andFe_supercell}. 

As expected, increasing the number of atoms in the structures in the training dataset led to better extrapolation to larger configurations. Consequently, we can anticipate that at some point, the model will be able to relax large structures without being pretrained on them. This is because the Agent will be capable of selecting the corresponding subspace for each atom and predicting its atomic shift based on the experience gained from relaxing smaller structures.

\subsubsection{Cumulative generalizability}

In this section, we assess the applicability of algorithms for more practical tasks. Specifically, we focus on training the model on simpler structures and subsequently use it to relax more complex structures, which are larger and have new types of chemical bonds.

\begin{figure}[ht] 
    \includegraphics[width=1\linewidth]{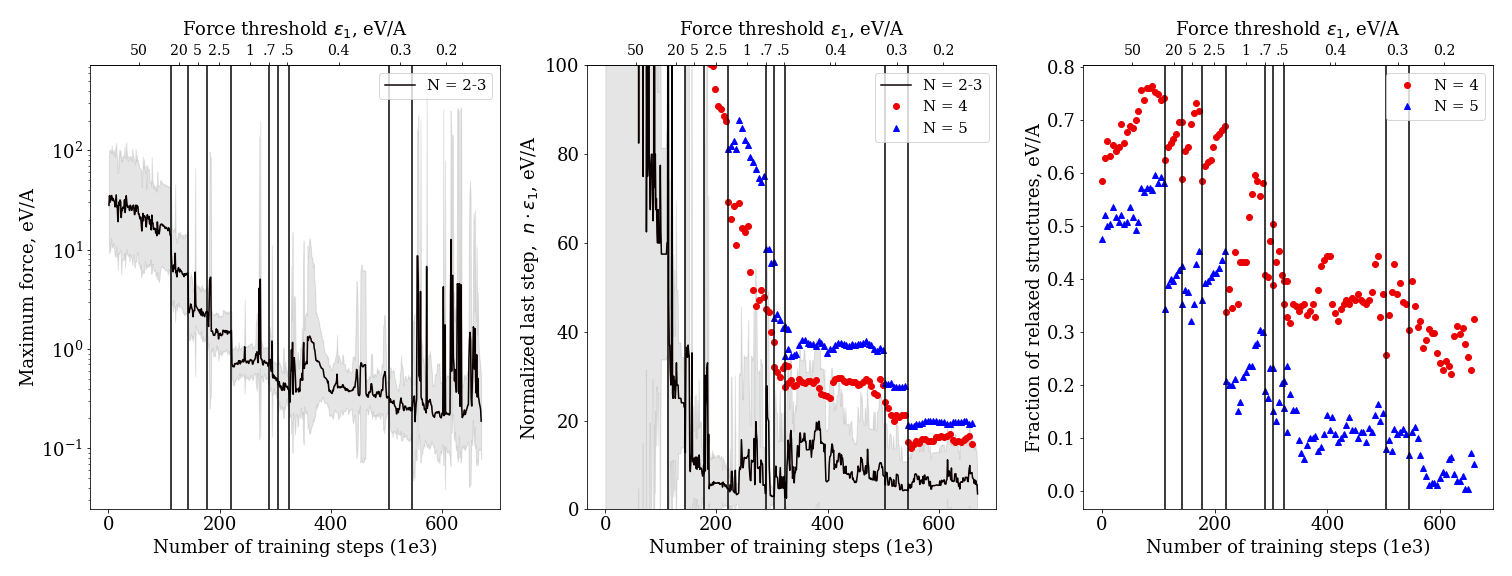}
    \caption{Learning curve of the TD3 model, trained on random Al and Fe configurations with $N = 2-3$ (black line) and validated on configurations with $N = 4$ (red dots) and $N = 5$ (blue triangles). We implemented a sequential decrease of the $\epsilon_1$ parameter, as described in Section \ref{Noise_and_gamma}, along with a \textit{hybrid} reward function. The corresponding $\epsilon_1$ values are shown on the upper axis. In every testing episode, the model was validated on datasets containing structures with $N = 4$ and $N = 5$ atoms per unit cell. For each structure, we performed 5 relaxation episodes, each starting with different initial distortions from the minimum. There were 50 structures for each $N$ (refer to Section \ref{Structures}).
    In the left plot, the maximum force at the last step, averaged over 20 relaxation episodes (1 per structure in the training dataset), is presented. The shaded area represents the standard deviation. The middle plot shows the normalized last step value $n\cdot\epsilon_1$ for both training and validation datasets, averaged over 20 relaxation episodes for the training dataset (1 per structure) and 250 episodes for each validation dataset (5 episodes per structure). The right plot illustrates the fraction of successful relaxation episodes out of the total number of relaxation episodes.} 
    \label{fig:Active}
\end{figure}

We use three datasets of Al-Fe structures with varying compositions generated using USPEX software. The first dataset consists of 20 structures, each containing 2-3 atoms per unit cell, and it is used to train the model. During the training process, we apply a sequential decrease of the $\epsilon_1$ parameter, as described in Section \ref{Noise_and_gamma}, along with a \textit{hybrid} reward function. The other two datasets, each comprising 50 structures with 4 and 5 atoms per unit cell, are used to validate the model's generalizability. In each testing episode, we relax these structures without incorporating them into the training dataset. The corresponding training curve is presented in Fig. \ref{fig:Active}.

To evaluate changes in the quality of the model's performance for different values of $\epsilon_1$, We use a concept of normalized last step, where the number of relaxation steps is multiplied by the force threshold to compensate for the increased difficulty as epsilon decreases. It allows to show the overall improvement of the relaxation algorithm and to compare the values between different $\epsilon_1$-regions. 

\begin{table*}
    \caption{Number of relaxation steps up to forces $0.2$ eV/A in the end of model training in Fig. \ref{fig:Active}. In the table mean values with standard deviation are presented. For $N=2-3$ the sample consisted of 20 relaxation episodes (1 per structure), for $N = 4$ and $N = 5$  it consisted of 250 episodes for each validation dataset (5 episodes per structure).}
	\centering 
    \begin{tabular}{ |c|c|c|c| } 
    \hline
     & $N = 2-3$ & $N = 4$ & Fig. $N = 5$\\\hline
     Relaxation steps  & 17.3 $\pm$ 5.8 &  73.1 $\pm$ 39.5 & 97.3 $\pm$ 12.8\\
    \hline
    \end{tabular}
    \label{tab:Last_step_ch_d}
\end{table*}

In the middle plot of Fig. \ref{fig:Active}, we observe that, on average, the normalized last step slightly changes after $\epsilon_1 = 0.3$ eV/A. This could be related to instability in the RL model, which is also reflected in the maximum force at the last step, as shown in the left plot of Fig. \ref{fig:Active}. The maximum force gradually decreases with the reduction of $\epsilon_1$, approaching $\epsilon_1$ as training progresses until it reaches $\epsilon_1 = 0.2$ eV/A. At this point, sharp peaks are observed, indicating possible instability due to a low sensitivity near the minimum discussed earlier. Because of this issue, switching to the \textit{step} reward function, as discussed in Section \ref{Noise_and_gamma}, did not improve the model, and thus, we do not provide the results for this reward function here. 

Regarding the model's generalizability, there is a strong correlation between the results for $N = 2-3$ and $N = 4-5$ as shown in Fig. \ref{fig:Active}. This indicates that our model can be applied to relax more complex structures without pre-training on them. However, the fraction of relaxed structures for reasonable values of $\epsilon_1$ (approximately 0.1 eV/A) ranges from 0.1 to 0.4, which is not a sufficient outcome.

There are two primary reasons for this. The first reason, discussed in Section \ref{Numbercompl}, is that the model needs to be trained on larger structures to efficiently identify the subspace of the most important neighbors for a given atom. The second reason is that not all interaction and symmetry patterns in structures with $4-5$ atoms per unit cell are present in structures with $N = 2-3$ atoms per unit cell. A possible solution would be to incorporate an active learning approach, sequentially adding structures with higher $N$ until the model is sufficiently trained to relax large structures. However, this approach is outside the scope of this study due to computational costs and may be considered in future research.

\section{Conclusions}\label{sec13}

 It has been demonstrated that the TD3 model, when trained on a single structure, performs comparably or even better than classical algorithms such as BFGS and CG, especially in the range of force convergence up to approximately \( \epsilon_1 \sim 10^{-1} - 10^{-2} \) eV/Å. However, achieving convergence to lower force values or training on several structures simultaneously remains challenging due to the sensitivity issues of the TFN model near minima. Despite this, relaxation to forces below $10^{-1} - 10^{-2}$ eV/Å is not critically problematic in practice. A hybrid approach can be employed: after reaching this force range, one can switch to classical algorithms, which can quickly reach a minimum since the potential energy surface (PES) in this region is close to a parabolic function. As a result, the current limitations discourage using this method as a universal optimizer. However, it is well-suited for tasks involving the relaxation of identical structures from different unstable initial states, such as in catalysis. In this case, the algorithm can be used as a baseline optimizer for commonly used systems.
 
We also explored the model's generalization capabilities from simple to complex structures. Initially, we assessed the model's performance as the number of atoms per unit cell increased. Results indicated that the model effectively relaxes supercells of similar structures, with performance improving as the size of the input structure in the training dataset increases. We anticipate that the model will eventually be capable of relaxing large structures without prior training on them, as the Agent will identify relevant subspaces for each atom and predict their atomic shifts based on experience gained from smaller structures. In addition, we found that the model can extrapolate its experience to cases where the complexity of chemical bonds increases. 

These findings illustrate the potential for crystal structure relaxation using deep reinforcement learning. Overcoming the mentioned challenges could lead to a universal structure optimizer that is faster for large configurations than traditional optimizers. Nonetheless, achieving this goal requires further development and additional computational resources for model training.

Our work underscores several critical aspects of reinforcement learning. Firstly, we highlight the importance of symmetry preservation in the action space for effective model training. We show that TFN model is more suitable to structure relaxation tasks compared to CGCNN, as it limits predicted actions to spaces with the same or higher symmetry than the input structure. Secondly, we point out the limitations of training with a piecewise-constant action-value function using a \textit{step} reward function. Direct optimization of the number of steps with this function is challenging, as the model may converge to a suboptimal policy due to a feedback loop of Critic and Actor convergence caused by the piecewise-constant nature of the true \(Q\)-function. However, pretraining the agent on smoother reward functions, such as \textit{force} or \textit{hybrid}, can mitigate this issue by increasing the likelihood of reaching terminal states early in training, a crucial condition for converging to an optimal policy.

Our study extensively examines the space of model hyperparameters and their impact on performance. We determined that simultaneous tuning of the discount factor and noise level enables the Agent to explore efficiently while balancing short- and long-term rewards to develop an optimal policy. Additionally, we investigated the influence of different reward functions, identifying the best way to combine them to maximize benefits across training epochs.

We compared various RL algorithms and found that TD3 is more efficient than SAC for structure relaxation tasks. However, this efficiency might be linked to exploration limitations in our SAC implementation. Future research will continue comparing different architectures and implementations of SAC and TD3 policies.

We hope our findings lay a foundation for further advancements in using reinforcement learning for structure relaxation tasks. 

\section*{Acknowledgements}

The calculations were performed on the Zhores cluster at Skolkovo Institute of Science and Technology. 

\section*{Declarations}

\begin{itemize}
\item Funding

This work was supported by the Russian Science Foundation (grant $\#$19-72-30043).

\item Conflict of interest/Competing interests

The authors declare no competing interests.

\item Data availability 

Cif files for structures generated with USPEX can be found and downloaded at: \url{https://github.com/ElenaTrukhan/RL_structure_relaxation/tree/main/structures}. 

\item Code availability 

The code that was used in the findings of this study is available from \url{https://github.com/ElenaTrukhan/RL_structure_relaxation.git}.

\item Author contribution 

E.T. implemented the code, trained the reinforcement learning (RL) models, analyzed the experimental results, and wrote the manuscript; E.M. contributed to the code implementation; A.O. and E.M. supervised the work; all authors discussed the results and reviewed the manuscript.
\end{itemize}

\begin{appendices}

\newpage

\section{Graph neural networks}\label{NodeEdge}

\subsection{Irreducible representations}
In this work, TFN models are implemented using the $e3nn$\cite{e3nn} Python library. Within the $e3nn$ framework, all input tensors are decomposed into the irreducible representations of the group O(3), which are labeled by the degree $l = 0,1,2,3, \ldots$ and parity $p \in (1,-1)$. When $p = 1$ the representation is called "even", otherwise, it is called "odd". For example, in general, $3\times3$ matrix $A$ can be decomposed into a scalar ($l = 0$, $p = 1$), pseudovector ($l = 1$, $p =1$) and even 2-nd order representations ($l = 2$, $p = 1$). In the $e3nn$ notation $A$ can be represented as $A = 1\times0e + 1\times1e + 1\times2e$.

\subsection{Convolutional operation}

The convolution operation in CGCNN is defined as follows \cite{Xie_2018}: 

\begin{equation}\label{Conv_CGCNN}
   v_i^{(t+1)} = v_i^{(t)} + \sum_{j,k}\sigma(z_{(i,j)_k}^{(t)}W_f^{(t)} + b_f^{(t)})\odot g(z_{(i,j)_k}^{(t)}W_s^{(t)}+b_s^{(t)}),
\end{equation}
where $z_{(i,j)_k}^{(t)} = v_i^{(t)} \oplus v_j^{(t)} \oplus u_{(i,j)_k}$, $\oplus$ denotes concatenation of vectors, $\odot$ denotes element-wise multiplication, $\sigma$ represents a sigmoid function, $W_f^{(t)}, W_s^{(t)}$ are the convolution weight matrices, $b_f^{(t)}, b_s^{(t)}$ are the  biases, $g$ is an activation function. 

In $e3nn$\cite{e3nn} the convolutional operation on node $a$ is implemented as the direct product of irreducible representations and can be written as:  

\begin{equation}\label{E3NN_filter}
   L_{acm_o}^{(l_o, p_o, l_f, p_f, l_i, p_i)}\left( \vec{r}_a, V_{acm_i}^{(l_i, p_i)}\right) = \sum_{m_f, m_i} C_{(l_i,m_i)(l_f, m_f)}^{(l_o,m_o)}
   \sum_{b \in S}\left(R(r_{ab})_{c}^{(l_f, l_i, p_f, p_i)} \right) Y_{m_f}^{(l_f)}(\hat{r}_{ab})V_{bcm_i}^{(l_i, p_i)},
\end{equation}
where $a$ and $b$ index the central atom being convolved and a neighboring atom respectively, $S$ represents graph nodes, $c$ corresponds to the channel. Each channel denotes a different instance of an irreducible representation. $V_{bcm_i}^{(l_i, p_i)}$ is the $m_i$-th component of the $b$-node feature in the $c$-th channel, which corresponds to the irreducible representation of O(3) group with degree $l_i$ and parity  $p_i$. $Y_{m_f}^{(l_f)}$  represents spherical harmonics of degree $l_f$ and order $m_f$. $\vec{r}_{a}$  represents the position of the $a$-th atom in the point  cloud, while $\vec{r}_{ab}$ denotes  the relative position from central atom $a$ to neighboring atom $b$, $r_{ab} = |\vec{r}_{ab}|$, $\hat{r}_{ab} = \vec{r}_{ab}/r_{ab}$. $C_{(l_i,m_i)(l_f, m_f)}^{(l_o,m_o)}$ are Clebsch-Gordan coefficients.

In Eq. (\ref{E3NN_filter}) $R(r_{ij})$ is a rotationally invariant radial function that contains all learnable weights. It is implemented as a multi-layer perception $R(r_{ij}) = W_n\sigma(....\sigma(W_2\sigma(W_1B(r_{ij})))$, where $B(r_{ij})$ denotes  the embedding of the interatomic distance of dimension $N_b$, $W_i$ are weight matrices and $\sigma(x)$ is the nonlinear function. In the given work the following embedding was used: 

\begin{gather}\label{embed}
\begin{aligned} 
   B(r) = \Biggl\{\exp  \left[ - \left( \frac{r - (a + s\cdot i)}{s}\right)^2 \right] \Biggl\}_{i=1}^N, 
\end{aligned}
\end{gather}
where $s = \frac{b-a}{N_b}$, $a$ and $b$ are the minimum and maximum value span by the basis respectively. 

\subsection{Graph construction}

During the conversion of crystal structures into crystal graphs, we use the following features for nodes and edges.

\subsubsection{Node features}
\begin{itemize}
    \item Scalar features: atomic number, standard atomic weight, Pauling electronegativity, atomic radius (pm), covalent radius (pm), first ionization energy (kJ/mol), electron affinity (kJ/mol), atomic volume (A$^3$), polarizability, number of electrons in the outermost shell, boiling point (K), specific heat capacity (J/gK).  All scalar features are normalized. In terms of $e3nn$ notation, these features correspond to $12\times0e$.  
    \item Vector features: force that acts on the atom, calculated with chosen potential (eV/A). As a vector, this feature corresponds to the $1\times1o$ irreducible representation. 
   
\end{itemize}

\subsubsection{Edge features}

Edge feature vector contained interatomic vectors $\vec{r}_{ij} = \vec{r}_i - \vec{r}_j$, which are then used for the embedding according to the rule in Eq. (\ref{embed}). In the given work $N_b$ =10, $a = 0$A, $b = 5$A. To determine the neighbors for each node, a sphere with a given radius  $r_{max} = 5$A and centered at the location of the atom is considered. 

\section{Twin delayed DDPG loss functions}\label{RL_ALG}

TD3 algorithm represents an enhanced version of Deep Deterministic Policy Gradient (DDPG) \cite{lillicrap2019continuous}. DDPG is based on the  Bellman optimality equation in Eq. (\ref{BellmanOpt}) and idea, that if optimal action-value function is know, optimal actions can be found as $a^{*}(s) = \text{argmax}_a Q^{*}(s,a)$ \cite{fujimoto2018addressing}.

So the loss function for $\pi_{\theta}(s)$ is obtained from the definition: 

\begin{equation}\label{TD3_pi_loss}
    L_{\theta}^{TD3} = -\mathbb{E}_{s \sim D} \Bigl[ Q_{\phi_1}(s,\pi_\theta(s)) \Bigr], 
\end{equation}

In Eq. (\ref{TD3_pi_loss}) one can observe the index in parameters $\phi$. The reason for this is that in TD3, two $Q$-functions, denoted as $Q_{\phi_1}$ and $Q_{\phi_2}$, are used to mitigate the overestimation of the real $Q$-function, that was observed in DDPG algorithm. Additionally, one may notice that the expected value, present in all the equations mentioned above, is replaced by the averaging over transitions $(s, s', a, d, r)$ sampled from a storage container $D$ referred to as the \textit{Replay buffer}. This buffer is of limited size and contains the experience gathered by the Agent during simulations.

The loss function for $Q_{\phi_i}$ is known as the \textit{mean-squared Bellman error} (MSBE), because it is derived from Eq. (\ref{BellmanOpt}) as follows:

\begin{equation}\label{TD3_Q_loss}
   L_{\phi_{i=1,2}}^{TD3} = \mathbb{E}_{(s,a,r,s',d) \sim D} \Bigl[ Q_{\phi_{i = 1,2}}(s,a) - \Bigl( r(s,a) + \gamma(1-d) \min_{i=1,2}Q_{\phi_{i, targ}}(s', \pi_{\theta_{targ}}(s')) \Bigr) \Bigr], 
\end{equation} 
where $Q_{\phi_{i, targ}}$ and $\pi_{\theta_{targ}}$ are target models. They are introduced to make target values in the loss function independent on the model parameters $\phi_{i=1,2}$ and $\theta$, thereby stabilizing the model update process. Unlike conventional models, the parameters of the target models do not get updated through gradient descent but change according to the following rule:  

\begin{equation}\label{TD3_targ}
    \omega_{targ} = \rho \omega_{targ} + (1-\rho)\omega, \; \omega = \{{\phi_1, \phi_2, \theta}\}
\end{equation}
where $\rho \in [0,1]$.

\section{Discussion of the step reward function}\label{step_discussion}

As it was mentioned in  Section \ref{Noise_and_gamma}, a straightforward use of the \textit{step} reward function does not lead to the better performance in terms of
the number of steps required for relaxation, which is to be expected according to the definition of this reward.

This can be partly explained by the findings of \cite{Sparsw_reward}, which investigates the failures of the TD3 algorithm in deterministic environments with sparse rewards, which is a type of reward that assigns a value of $0$ for all transitions except terminal ones. In such cases the TD3 Agent tends to converge to a suboptimal policy if it does not encounter any non-zero rewards at the beginning of training because of the piecewise-constant nature of the true action-value function $Q^{\pi}$ corresponding to a poor policy which leads to gradients $\nabla_\theta Q(s, \pi_\theta(s)) \simeq 0$ and causes a closed loop of zero updates for both the Critic and the Actor. 

While the reward function \textit{step} is not sparse, it is intuitively clear (see Section \ref{RL_notions}) that its action-value function is also piecewise-constant. To demostrate that let's consider a "good" policy $\pi$, which, starting in the given state $s_0$ and taking action $a_0$, reaches the terminal state in $n$ steps ("good" in comparison to the policy that never allows the Agent to achieve the terminal state). Its $Q$-function is given by:

\begin{equation}\label{Q_step}
    Q^{\pi}(s_0, a_0) = R(\tau) = \sum_{t = 0}^ {n-1} (-1)\cdot\gamma^t + 0= -\frac{\gamma^n - 1}{\gamma - 1}
\end{equation}
where $\gamma$ is the discount factor. 

We can see that for a "good" policy, the function domain of $Q^{\pi}(s_0, a_0)$ is partitioned into regions with identical values, as determined by Eq. (\ref{Q_step}). Consequently, even for a "good" policy, the converged action-value function remains piecewise-constant, and it is true for both noise settings. The crucial distinction between them lies in the fact that adding noise to action diminishes the probability of reaching the terminal state during training by disrupting the symmetry of actions and removing them from the space of actions relevant for the given structure, as discussed earlier. This aligns with findings in \cite{Sparsw_reward}, where delayed discovery of the initial reward often leads to the learning process becoming stuck in a sub-optimal policy due to the closed loop problem described above. This scenario applies to our task, as the absence of encounters with the terminal state may lead the Agent to consider the sum in Eq. (\ref{Q_step}) for each action-state $(s,a)$ pair as infinite, resulting in a constant value of $Q^{\pi}(s, a) = \frac{1}{\gamma - 1}$ for all action-state pairs, resembling the situation described in \cite{Sparsw_reward}. As a result, it is crucial to provide the Agent with exhausting experience in the vicinity of the terminal step in the beginning of training before convergence of the models.  However, this explanation remains unverified in our study but it could be a subject of investigation in the future. 

\section{Sensitivity}\label{Sens}

\begin{figure}[htbp]
\center{\includegraphics[width=1\linewidth]{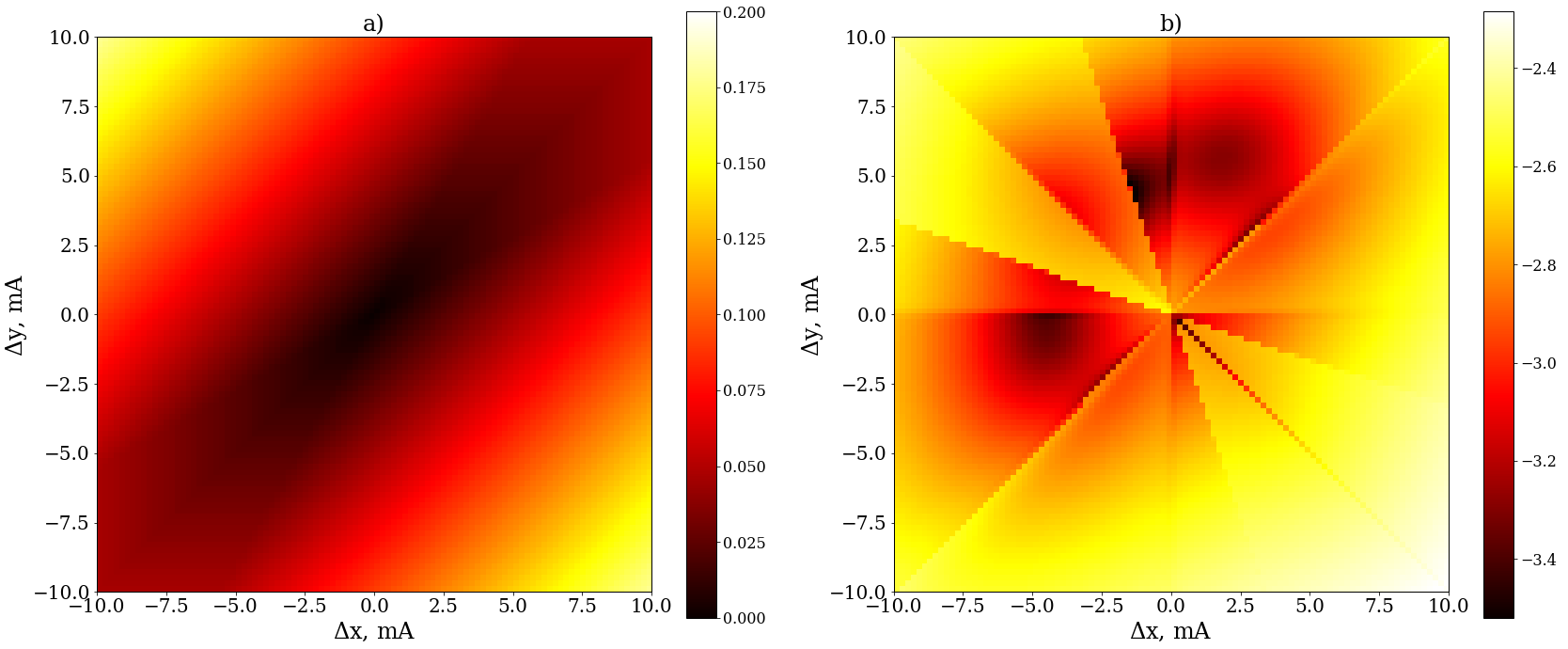}} \\
\caption{a) Maximum norm of the force $\max_{n\in [1,N]}|\vec{f}_n(s^*)|$ in the structure $s^*$, distorted from the local minimum $s_0$ by shifting the first atom in the structure to the vector  $\{\Delta x,\Delta y,0\}$. The force is in eV/A; b) Logarithm of the norm of the difference in the model's predictions for the first atom, $lg(|\Delta \vec{r}_1(s^*) - \Delta \vec{r}_1(s_0)|)$, as a function  of $\Delta x$ and $\Delta y$. The results are measured for the structure $\#15$ in the Table \ref{Table_struct} using the model, trained to perform the relaxion up to forces $0.2$ eV/A.}

\label{fig:Sens10}
\end{figure}

To investigate the sensitivity of our model near the minimum, we measured how the predictions differ between the structure corresponding to the minimum, $s_0$, and the slightly distorted structure, $s^*$. We introduced distortions by shifting the first atom from the minimum in $X-Y$ plane: $\vec{r}_1(s^*) = \vec{r}_1(s_0) + \{\Delta x,\Delta y,0\}$. For this experiment, we selected a structure with 10 atoms (\#15 in the Table \ref{Table_struct}).  As mentioned in Section \ref{Numbercompl}, we were able to train our model to relax this structure only up to forces of $0.2$ eV/A, while further relaxation was challenging. Therefore, for predicting shifts $\Delta \vec{r}_1$ , we used the model trained to relax up to $0.2$ eV/A and considered the range of $\Delta x$ and $\Delta y$, where $\max_{n\in [1,N]}|\vec{f}_n(s^*)| \leq 0.2$ eV/A (see Fig.\ref{fig:Sens10}a). In Fig. \ref{fig:Sens10}b) one can observe the logarithm of the norm of the difference in the model's predictions for the first atom, $lg(|\Delta \vec{r}_1(s^*) - \Delta \vec{r}_1(s_0)|)$, as a function  of $\Delta x$ and $\Delta y$.  

In the close vicinity of the minimum, all input data in the crystal graphs change smoothly. However, in Fig.  \ref{fig:Sens10}b), a pronounced discontinuity in the model's predictions can be observed. This indicates that the model struggles to accurately predict atomic shifts further from the minimum.

\section{Exploration in RL}\label{Ap_noise_to_action} 

Section \ref{Comp_archs} demonstrated that reducing the action space in accordance with the task's symmetry enables the Agent to identify an optimal policy more quickly. We also observed a similar effect in relation to exploration strategy.

In this study, we implemented an alternative approach to exploration for TD3 algorithm by \textit{adding noise to action}. In this case, small uniformly distributed noise is added independently to each coordinate, $(\Delta \vec{r}_i)_j \rightarrow (\Delta \vec{r}_i)_j + \xi_{ij}$, where $\xi_{ij} \sim U(-\lambda, \lambda)$, and $\lambda$ represents the \textit{noise level}.

\begin{figure}
\center{\includegraphics[width=1\linewidth]{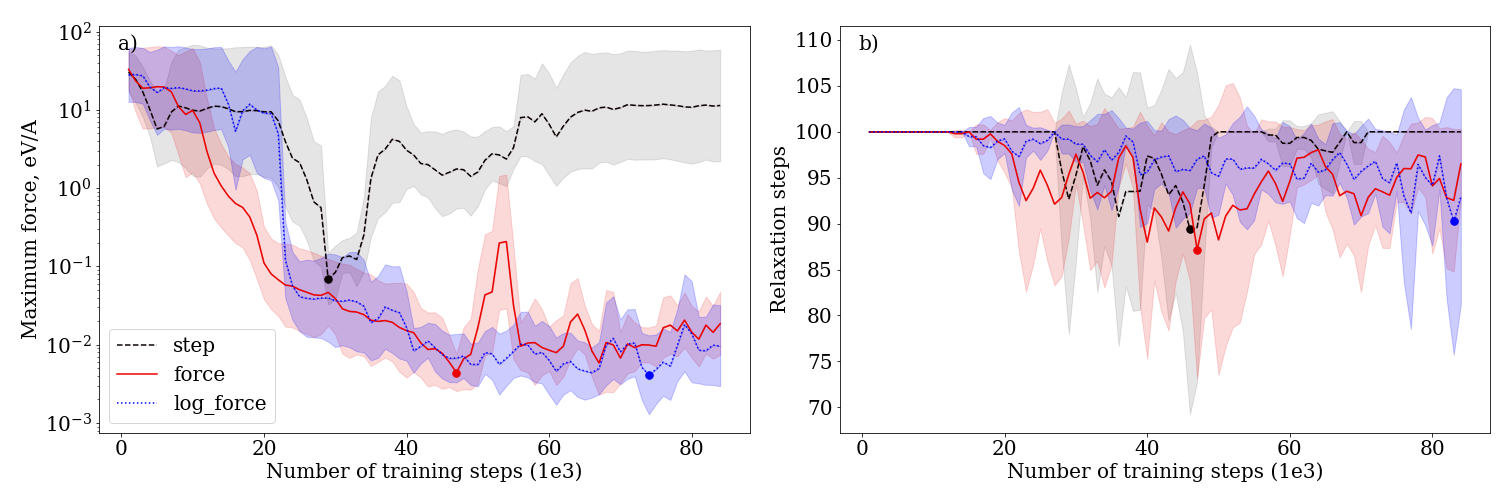}} \\

\caption{Learning curves of TD3 Agents with different reward functions, given by Eq. (\ref{R1_atom})-(\ref{R3_atom}) with \textit{adding noise to action} exploration stratagy. The Agents were trained on CsCl-type structure of AlFe. Curves are averaged over 5 seeds, with the shaded area representing the standard deviation. Circles mark the lowest average last step and maximum force.}

\label{fig:Different_rewards_action}
\end{figure}

To evaluate this approach, we compared it with the \textit{adding noise to state} method by training the TD3 Agent with different noise settings on the CsCl-type structure of AlFe. The results for \textit{adding noise to action} are shown in Fig. \ref{fig:Different_rewards_action} and the comparison is summarized in Table \ref{tab:different_rewards}. 

\begin{table}[!htb]
    \centering
     \caption{
     The best results achieved by the TD3 Agents with different reward functions and noise settings. \textit{Average}  indicates the lowest maximum force and relaxation steps from the averaged learning curves, marked by circles in Fig. \ref{fig:Different_rewards}a-b) and Fig. \ref{fig:Different_rewards_action}. \textit{The best results over all trials} represents the top-performing outcomes across all seeds used for averaging. The lowest-performing maximum force and relaxation steps are highlighted.}
    \begin{tabular}{ | c | c | c | c | c | c | c |  }
     \hline
     & \multicolumn{3}{|c|}{Noise to state} & \multicolumn{3}{|c|}{Noise to action} \\
     \hline
     & force & step & log force & force & step & log force\\
     \hline
     \multicolumn{7}{|c|}{Average} \\ 
     \hline
     Relaxation steps & 86 $\pm$ 25 & \colorbox{Goldenrod}{79 $\pm$ 23} & \colorbox{Goldenrod}{79 $\pm$ 22} & 85 $\pm$ 16 & 89 $\pm$ 20 & 88 $\pm$ 18 \\
     Force, meV/A  &  8.9 $\pm$ 12.5 & 1.6 $\pm$ 1 & \colorbox{Goldenrod}{1.1 $\pm$ 0.7}  & 2.8 $\pm$ 1.5 & 63.5 $\pm$ 50.2 & 3.5 $\pm$ 2.2 \\
     \hline
     \multicolumn{7}{|c|}{The best results over all trials} \\ 
     \hline 
     Relaxation steps  & \colorbox{Goldenrod}{37.2} & 44 & 44.9  & 57.2 & 47.4 & 51.9 \\
     Force, meV/A  &  0.043  & 0.054   & \colorbox{Goldenrod}{0.042} & 0.06 & 0.048 & 0.09 \\
     \hline
    \end{tabular}
    \label{tab:different_rewards}
\end{table}

The comparison clearly shows that, for all reward functions, \textit{adding noise to state} consistently outperforms \textit{adding noise to action}. This highlights the importance of preserving symmetry in RL tasks, as previously discussed. Adding noise to output vectors disrupts the symmetry encoded in the TFN model. As a result, this exploration strategy fails to provide relevant experiences for the Agent, because, firstly, due to Curie’s principle, the model cannot replicate such actions, and secondly, these actions may not even fall within the subspace of atomic shifts relevant to the structure due to its symmetry. Conversely, \textit{adding noise to state}, as implemented in this study, successfully avoids these issues.

The most pronounced difference is observed with the \textit{step} reward function due to the convergence issues discussed in Section \ref{Noise_and_gamma} and Appendix \ref{step_discussion}. Adding noise to actions reduces the probability of reaching the minimum using the initial suboptimal policy and leads to convergence loops, as described in Appendix \ref{step_discussion}.

These findings further underscore the advantage of restricting the possible actions to align with the task's symmetry, thereby enhancing model performance.

\section{Additional greedy exploration}\label{Ap_greedy}

\begin{figure}[htbp]
\begin{minipage}[h]{0.48\linewidth}
\center{\includegraphics[width=1\linewidth]{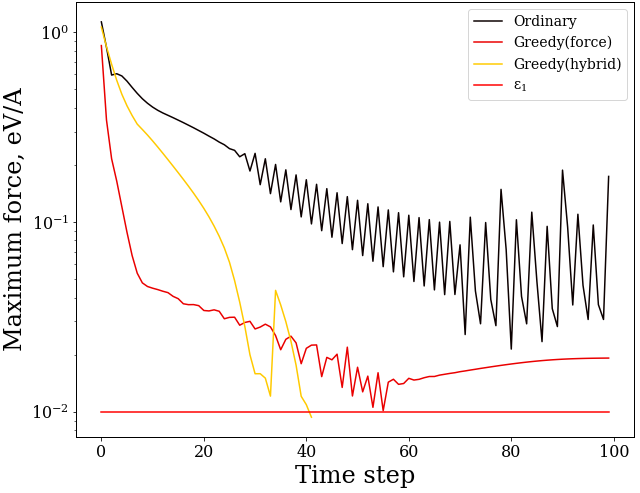}} a) \\
\end{minipage}
\hfill
\begin{minipage}[h]{0.48\linewidth}
\center{\includegraphics[width=1\linewidth]{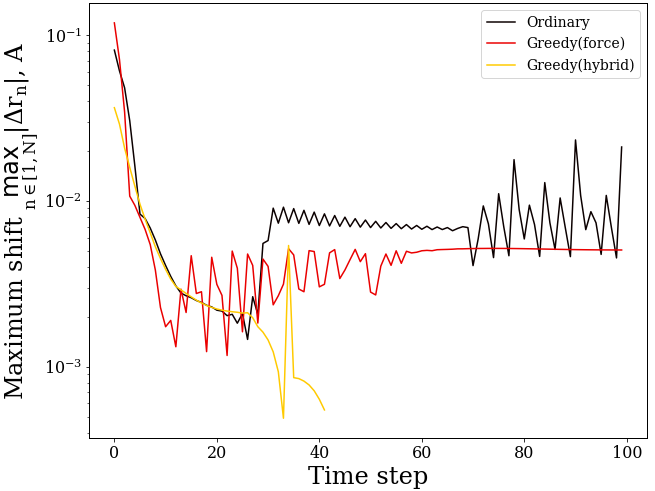}} b) \\
\end{minipage}
\caption{Several relaxation trajectories, conducted by the Agent trained on hypothetical I4/mmm structure of Al, are depicted. Different colors correspond to different initial states $s_0$ from which relaxation begins.  The dependency of the a) maximum force $\max_{n\in [1,N]}|\vec{f}_n(s_t')|$ b) maximum shift $\max_{n\in [1,N]}|\Delta\vec{r}_n|$ on the time step is illustrated. Line color denotes the model with which relaxation was carried out: 1) black corresponds to the ordinary model; 2) red corresponds to the greedy model trained with \textit{force} reward function; 3) yellow corresponds to the greedy model trained with \textit{hybrid} reward function.}
\label{fig:Stuck}
\end{figure}

\begin{figure}[htbp] 
    \includegraphics[width=\linewidth]{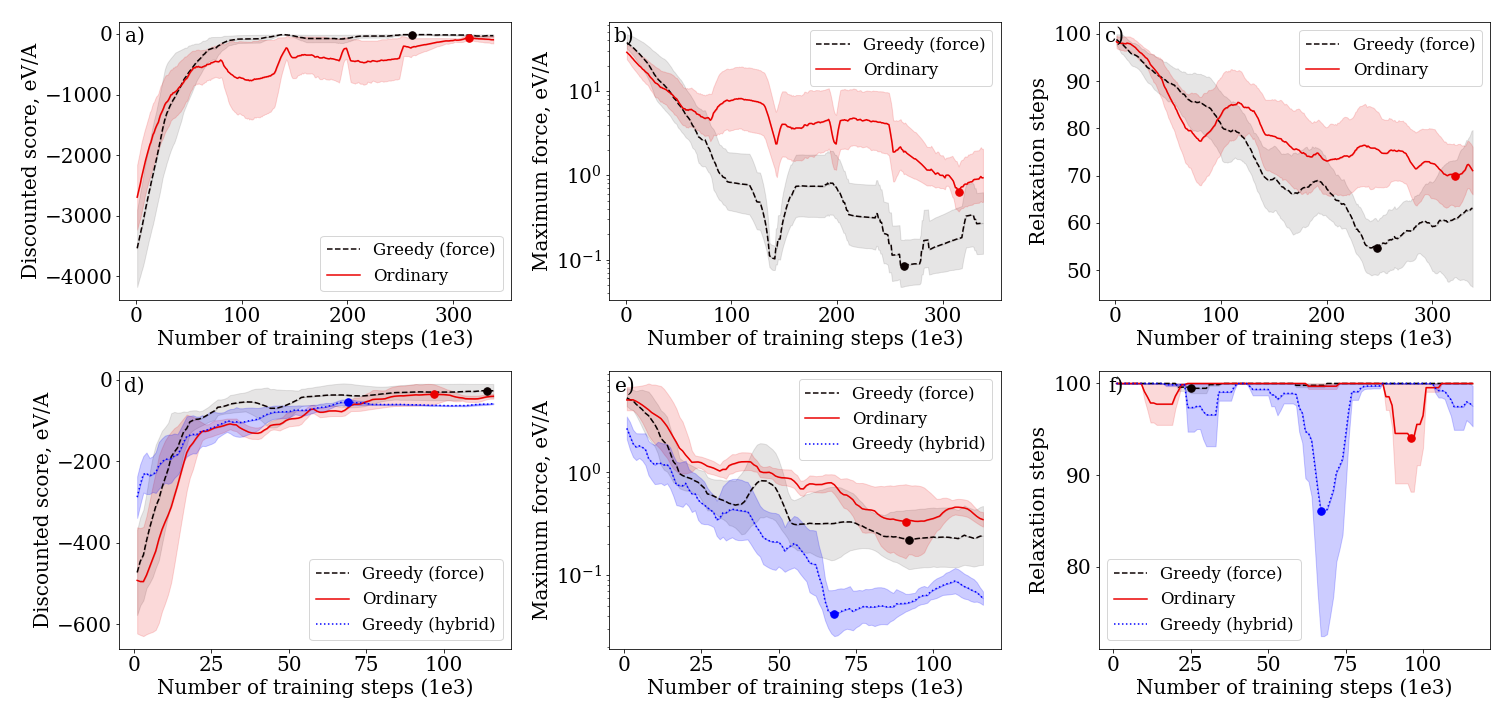}
    \caption{Comparison of TD3 Agents trained with ("Greedy (force)" and "Greedy (hybrid)") and without ("Ordinary") greedy exploration on a-c) the set of Al and Fe structures listed in Table \ref{Table_struct} d-f) hypothetical I4/mmm structure of Al with 4 atoms per unit cell. The bold lines represent the learning curve averaged over 2 trials with different random seeds and shaded regions show corresponding standard deviation.}
    \label{fig:Sev_str_greedy_vs_norm}
\end{figure}

The performance of the TD3 Agent, with and without additional greedy exploration, was evaluated using different reward functions in two cases: (1) the hypothetical I4/mmm structure of Al and (2) a dataset of monoatomic Al and Fe structures (see Fig. \ref{fig:Sev_str_greedy_vs_norm}). Fig. \ref{fig:Stuck} illustrates how the relaxation trajectory varies for the same initial configuration of the hypothetical I4/mmm structure of Al when using different trained models.

The implementation of additional greedy exploration improves the model's performance, by reducing both the maximum force on the final step and the number of relaxation steps.

\section{Soft Actor Critic}\label{SAC_sec}

In Soft-Actor Critic (SAC) \cite{haarnoja2018soft}\cite{OpenAI} the policy is stochastic so the entropy regularization term is introduced to the objective to provide with an exhaustive exploration of the Environment:  

\begin{equation}\label{Q_SAC_Bellman}
    Q_H^{\pi^*}(s, a) = \mathbb{E}_{s' \sim P(\tau|\pi)} \left[ R(s, s', a) + \gamma \max_{a'} \left( Q_H^{\pi^*}(s', a') - \alpha \log \pi(a'|s')  \right) \right], 
\end{equation}  
where $\alpha$ is the trade-off coefficient. It determines the relative importance of the entropy term $H(\pi) = \text{E}_{a\sim \pi} \left[ - \log \pi(a|s) \right]$ against the reward and thus controls the stochasticity of the optimal policy. 

As well as TD3, SAC utilizes two types of neural networks for \textit{Actor} and \textit{Critic}.

\subsection{Loss functions}  

Soft Actor-Critic (SAC) is an off-policy algorithm that optimizes a stochastic policy, effectively bridging the gap between stochastic policy optimization and DDPG-style methods. Entropy regularization leads to a modification of the objective function $ J(\pi) = \mathbb{E}_{\tau \sim \pi,f} \left[ \sum_{t=0}^T \gamma^t \left(r_t + \alpha H(\pi(\cdot|s_t)) \right)\right]$. $Q^{\pi}$ and $V^{\pi}$ are also adjusted correspondingly \cite{haarnoja2018soft}. 

SAC, similar to TD3, employs the concept of deriving an optimal policy from an optimal $Q$-function. There are also two models $Q_{\phi_1}$ and $Q_{\phi_2}$, used for unbiased approximation of $Q^*$, an optimal policy model  $\pi_\theta$, and target models for the $Q$-functions. The loss function for $Q_{\phi_i}$is akin to the MSBE in Eq. (\ref{TD3_Q_loss}), but with additional entropy contribution:  

\begin{equation}\label{SAC_Q_loss}
   L_{\phi_{i=1,2}}^{SAC} = \mathbb{E}_{(s,a,r,s',d) \sim D} \Bigl[ Q_{\phi_{i = 1,2}}(s,a) - \Bigl( r(s,a) + \gamma(1-d) \min_{i=1,2}Q_{\phi_{i, targ}}(s', a') - \alpha \log\pi_{\theta}(a'|s')\Bigr) \Bigr],     
\end{equation} 
where $a' \sim \pi_{\theta}(\cdot|s')$. 

Since the deep model cannot directly predict the probability density function, it is assumed that $\pi_\theta \sim N(\mu_\theta, \sigma_\theta)$. Consequently, the model approximates the expectation value $\mu_\theta$ and variance $\sigma_\theta$ of the normal distribution. In this scenario, the loss for the policy takes the following form: 

\begin{gather}\label{SAC_pi_loss}
\begin{aligned} 
   L_{\theta}^{SAC} = -\mathbb{E}_{s \sim D, \xi \sim N(0, I)} \Bigl[ \min_{i=1,2}Q_{\phi_i}(s,a_\theta(s,\xi)) -\alpha \log\pi_\theta(a_\theta(s,\xi)|s) \Bigr], 
\end{aligned}
\end{gather}
where $a_\theta(s,\xi) = \mu_\theta(s) + \sigma_\theta \odot \xi$, $\xi \sim N(0, I)$. 

\subsubsection{Implementation of exploration}\label{SAC_policy}
In SAC exploration is inherently incorporated through entropy regularization. The policy is stochastic and given by the probability distribution $\pi(a|s)= P(a|s)$, $a_t  \sim \pi(\cdot|s_t)$. Since the deep model cannot directly predict PDF, it is assumed that $\pi_\theta \sim N(\mu_\theta, \sigma_\theta)$, where in a general case $\mu_\theta$ is $k$-dimensional mean vector and $\sigma_\theta$ is $k \times k$ covariance matrix. Consequently, the model approximates $\mu_\theta$ and $\sigma_\theta$.  

However, in the given task a challenge arises because $a_t$ is a graph, containing atomic shifts. Approximating the optimal policy with a set of independent normal distributions for each vector would hardly provide sufficient Agent performance, as these values are connected by various relationships, such as E(3)-equivariance or symmetry constraints of the structure.

The solution presented in this work is to introduce stochasticity solely to the norm of the vectors. In this approach, the model predicts a graph where each node contains a vector $\vec{\mu}_i$ and a scalar $\sigma_i$. Then expectation values $|\vec{\mu}_i|$ and unit direction vectors $\vec{n}_i = \vec{\mu}_i/|\vec{\mu}_i|$ are computed, normal distribution $N(|\vec{\mu}_i|, \sigma_i)$ is constructed for each node, and the magnitude $|\Delta \vec{r}_i|$ is sampled as $|\Delta \vec{r}_i| \sim N(|\vec{\mu}_i|, \sigma_i)$. The atomic shift for the $i$-th atom is then obtained as $\Delta \vec{r}_i = |\Delta \vec{r}_i|\cdot \vec{n}_i$. This approach ensures compliance with the norm distribution constraint under rotation and preserves the correlations between shift directions detected by the model. However, it restricts the Agent's exploration capabilities, as stochasticity is only introduced to the shift lengths.

\subsubsection{Hyperparameters exploration and comparison with TD3}\label{SAC_policy_expl}

Similarly to TD3 we investigated the influence of the discount factor and exploration level on the performance model, using CsCl-type structure of AlFe, but the main difference is that in the case of SAC exploration is controlled by the trade-off parameter $\alpha$. Results of the tests are presented in Fig. \ref{fig:Different_gamma_and_noise_SAC_TD3}b) and more detailed comparison of learning curves is also shown in Fig. \ref{fig:Different_gamma_and_alpha_SAC}. 

As it can be seen, for SAC the best performance correspond to the model trained with the smallest $\gamma$ factor and $\alpha$, which proves the assumption, derived in Appendix \ref{Disc_f_and_noise} that a lower discount factor with small exploration allows the model to concentrate on the immediate slightly noised outcome and build in this way efficient policy.   

As for comparison of TD3 and SAC, it can be seen over the range of hyperparameters that TD3 outperforms SAC in number of steps required for relaxation, so that is the reason why TD3 is used mainly in the given work. The possible reason might be that policy implementation, described in Section \ref{SAC_policy}, does not provide the Agent with exhaustive exploration, because it is limited only to stochasticity in the norm of atomic shifts $\Delta \vec{r}_i$, but does not cover their directions. So further improvement of the model is needed to increase the quality of the algorithm.  

 \section{Discount factor and level of exploration}\label{Disc_f_and_noise}

\begin{figure} 
	\includegraphics[width=\linewidth]{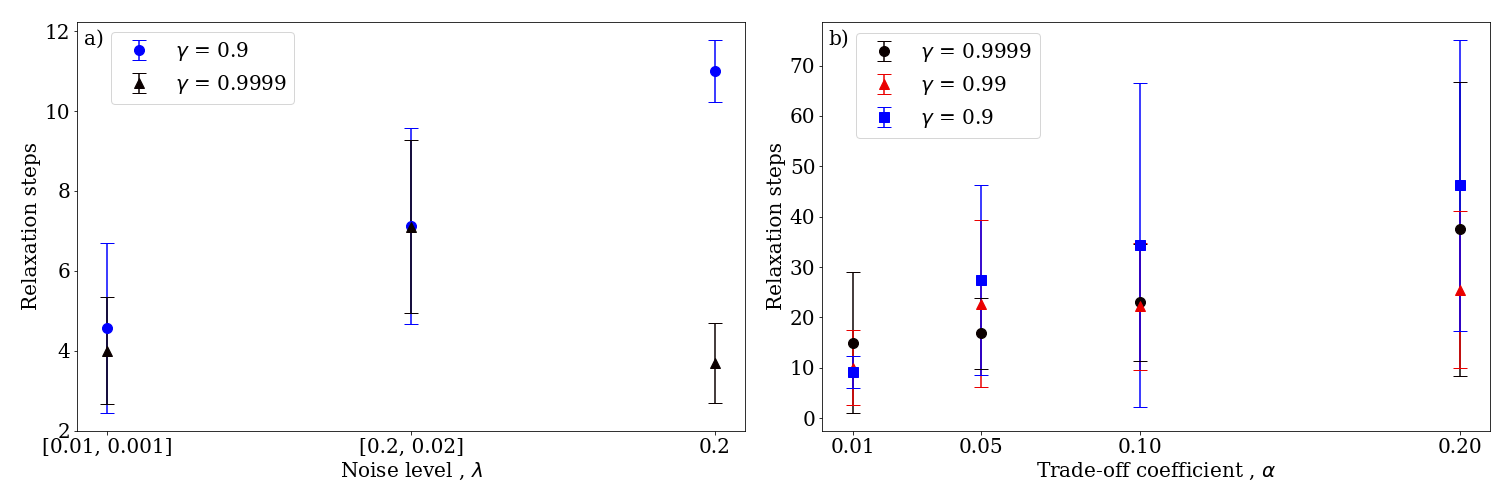}
\caption{Performance of the best in terms of relaxation steps of TD3 and SAC models (marked with circles in Fig. \ref{fig:Different_gamma}b,e,h) and Fig. \ref{fig:Different_gamma_and_alpha_SAC}b,e,h). The Agents were trained with different discount factors $\gamma$ and level of exploration on CsCl-type structure of AlFe. The figure shows the mean value with standard deviation averaged over 10 relaxation test episodes.}
\label{fig:Different_gamma_and_noise_SAC_TD3}
\end{figure}

The discount factor $\gamma$ is a crucial parameter that influences how much the Agent values future rewards compared to immediate rewards. A higher $\gamma$ assigns greater significance to long-term rewards, prompting the Agent to carefully consider future consequences in its decision-making process. However, depending on the task, the relationship between episodes may weaken as the number of time steps between them increases. This weak correlation, combined with the inclusion of long-term rewards, can introduce additional uncertainty into the estimation of value functions. In this case a lower $\gamma$ is more suitable. Small discount factor also motivates the Agent to take more risks. Thus, both small and large $\gamma$ have positive and negative effects, so in this work, we examined the performance of the model with different $\gamma$  and levels of exploration. In the case of TD3 exploration is controlled by the parameter $\lambda$.

The lowest number of relaxation steps obtained in this work is presented in Fig. \ref{fig:Different_gamma_and_noise_SAC_TD3}a). More detailed comparison of learning curves is also shown in Fig. \ref{fig:Different_gamma}. 

As it can be seen, for TD3 in scenarios with high noise levels, increasing the discount factor leads to more stable convergence and improved performance in terms of score, last step, and maximum force. Conversely, in low-noise situations, a lower discount factor is preferable. This discrepancy arises from the impact of noise on value estimates. While high noise enhances exploration efficiency, it also introduces variance in value estimates. Consequently, a higher discount factor stabilizes policies by prioritizing long-term rewards over noisy immediate feedback. However, in low-noise scenarios, rewards are mainly generated by the policy itself, and a lower discount factor allows the Agent to focus more on the immediate results obtained by its policy.

\section{Additional figures and Tables}\label{Ap3}

\subsection{Hyperparameters tuning: discount factor and level of exploration }

\begin{figure}[h]
\includegraphics[width=1\linewidth]{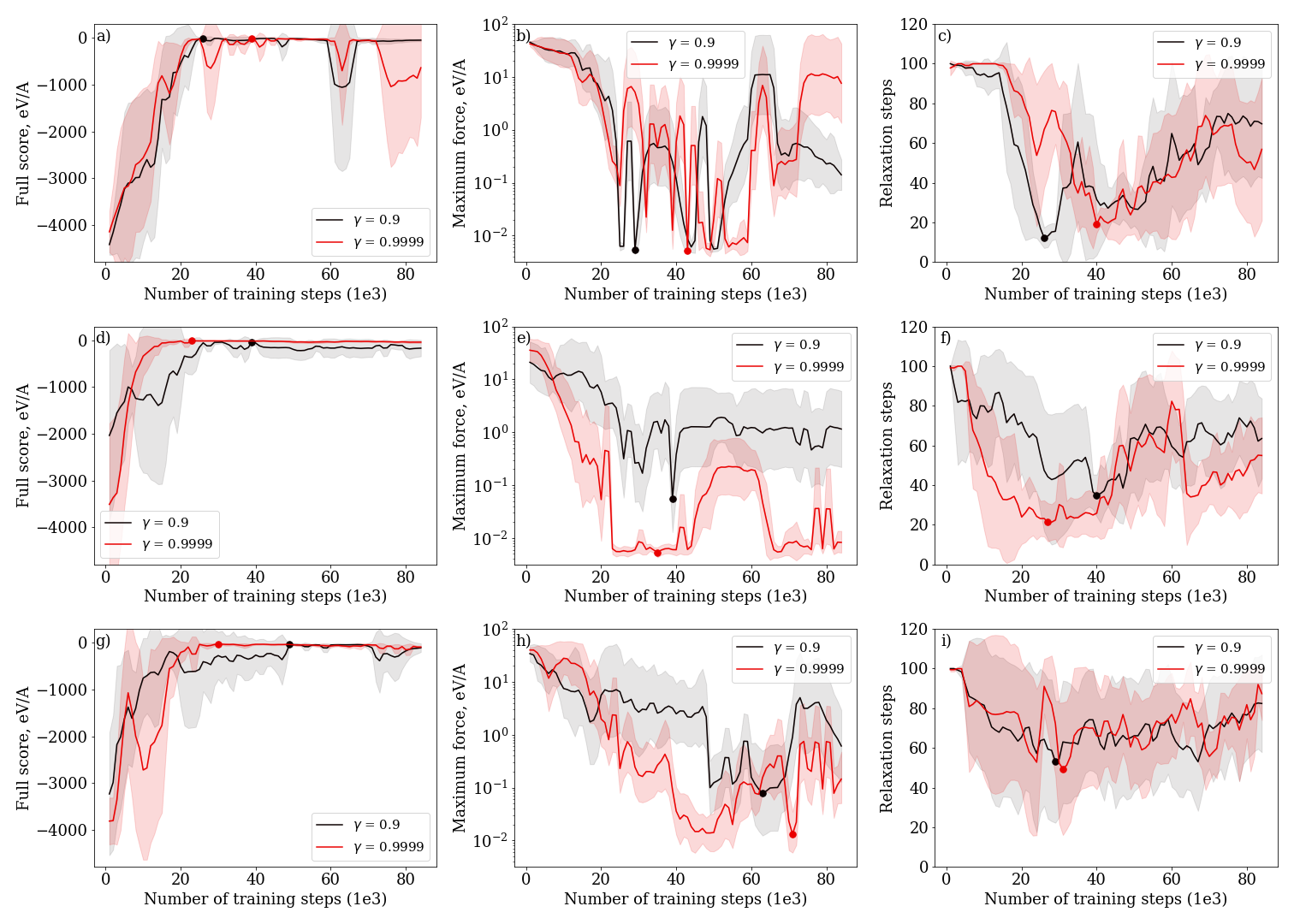}
\caption{Learning curves comparing the performance of TD3 Agents with different discount factors $\gamma$ and noise levels $\lambda$: a-c) $\lambda$ = [0.01, 0.001], d-f) $\lambda = 0.2$, g-i) $\lambda = [0.2,0.02]$. Curves are averaged over 5 seeds, with the shaded area representing the standard deviation across seeds. Circles mark the lowest average number of relaxation steps and maximum force, as well as the highest score.}
\label{fig:Different_gamma}
\end{figure}

\begin{figure}[h]
\center{\includegraphics[width=1\linewidth]{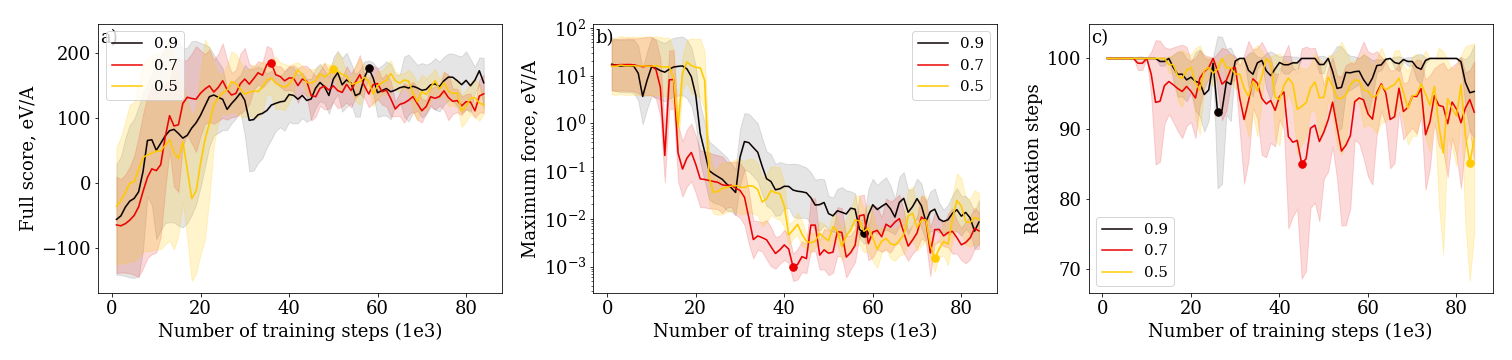}}\\
\caption{Learning curves comparing the performance of TD3 Agents with different discount factors $\gamma$. The Agents were trained on CsCl-type structure of AlFe  with a \textit{log force} reward function and a distortion parameter $\beta = 0.5$. Curves are averaged over 5 seeds, with the shaded area representing the standard deviation across seeds. Circles mark the lowest average number of relaxation steps and maximum force, as well as the highest score.}
\label{fig:Different_gamma_log_force}
\end{figure}

\begin{figure}[htbp]
\center{\includegraphics[width=1\linewidth]{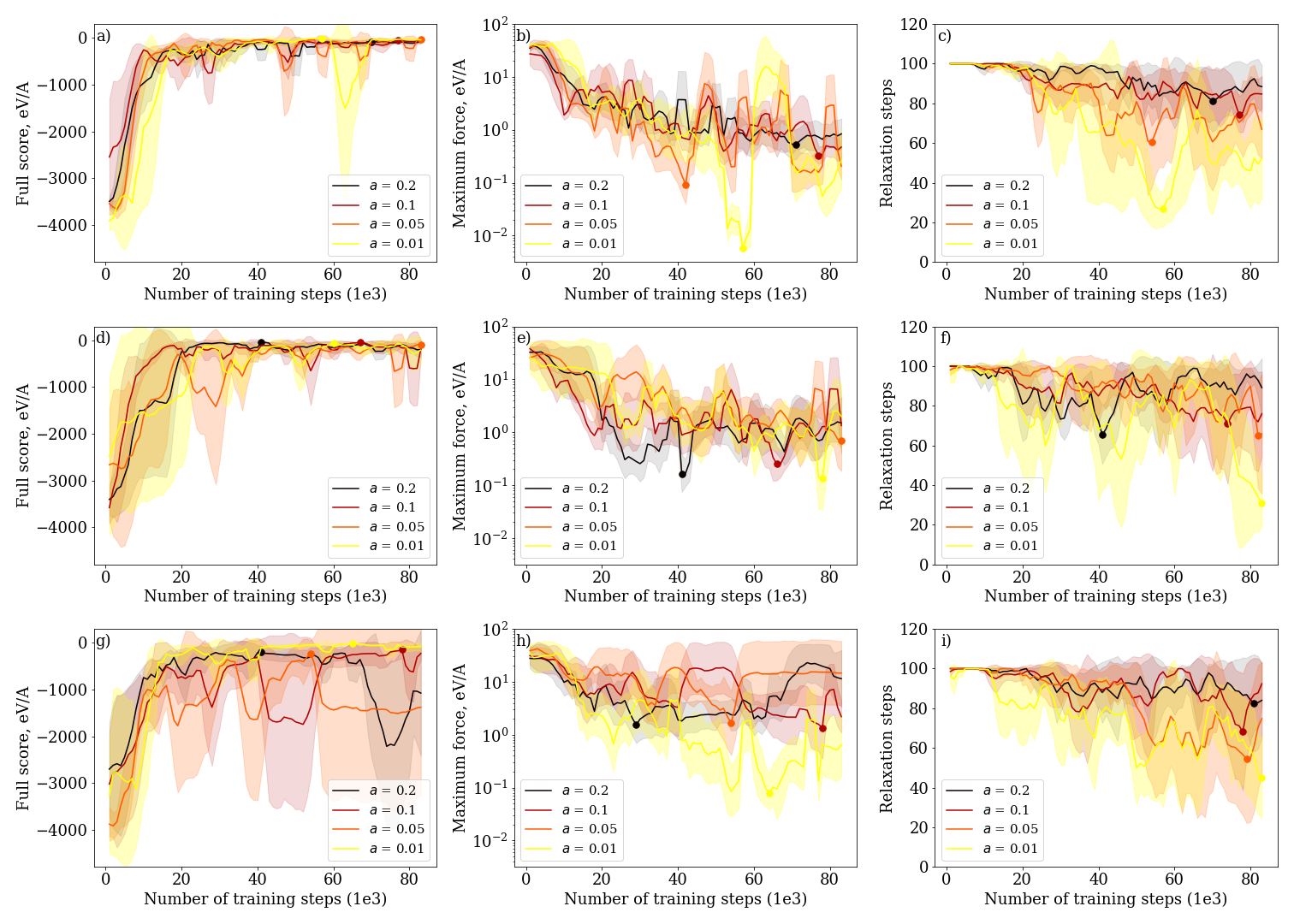}}\\
\caption{Learning curves comparing the performance of SAC Agents with different discount factors $\gamma$ and entropy trade-off coefficients $\alpha$: a-c) $\gamma$ = 0.9, d-f) $\gamma = 0.99$, g-i) $\gamma = 0.9999$. The Agents were trained on CsCl-type structure of AlFe with \textit{force} reward function and a distortion parameter of $\beta = 0.5$. Curves are averaged over 3 seeds, with the shaded area representing the standard deviation across seeds. Circles mark the lowest average number of relaxation steps and maximum force, as well as the highest score.}
\label{fig:Different_gamma_and_alpha_SAC}
\end{figure}

Fig. \ref{fig:Different_gamma}, Fig. \ref{fig:Different_gamma_log_force}, Fig. \ref{fig:Different_gamma_and_alpha_SAC} and Table \ref{tab:Different_gamma}, Table \ref{tab:Different_gamma_and_alpha_SAC}.

\begin{table}[htbp]
    \centering
     \caption{The best results of the TD3 Agents with different discount factors $\gamma$ and noise levels $\lambda$. \textit{Average} corresponds to the highest score and the lowest maximum force and number of relaxation steps of the averaged learning curves, marked with circles in Fig. \ref{fig:Different_gamma}. \textit{Best} represents the top-performing results among all seeds used for averaging. The highest scores and the lowest maximum force and number of number of relaxation steps are highlighted.}
    \begin{tabular}{ | c | c | c | c | c | c | c | c |  }
     \hline
      \multicolumn{2}{|c|}{} & \multicolumn{2}{|c|}{Score, eV/A} & \multicolumn{2}{|c|}{Relaxation  steps} & \multicolumn{2}{|c|}{Max force, meV/A} \\
     \hline
       $\lambda$ & $\gamma$ & Average & Best & Average & Best & Average &  Best \\
    \hline
    \multirow{ 2}{*}{[0.01, 0.001]} & 0.9 & \colorbox{Goldenrod}{-3.4 $\pm$ 0.5} & -1.0 & \colorbox{Goldenrod}{11 $\pm$ 4} & 4.3 & \colorbox{Goldenrod}{4.7 $\pm$ 1.1} & 3.2 \\
    & 0.9999 & -7.8 $\pm$ 4.4 & \colorbox{Goldenrod}{-0.5} & 14.5 $\pm$ 5.5 & 4.0 & 5 $\pm$ 0.6 & 2.5 \\
    \hline
    \multirow{ 2}{*}{0.2} & 0.9 & -24.7 $\pm$ 8.7 & -3.2 & 30 $\pm$ 9 & 11 & 6.2 $\pm$ 1.3 & 3 \\
    & 0.9999 & -6.8 $\pm$ 5.1 & -1.2 & 20.4 $\pm$ 10.1 & \colorbox{Goldenrod}{3.7} & 4.8 $\pm$ 0.9 & 2.7 \\        
    \hline
    \multirow{ 2}{*}{[0.2, 0.02]} & 0.9 & -32.5 $\pm$ 19.3 & -1.5 & 48 $\pm$ 26 & 5.5 & 65.6 $\pm$ 118.7 &  2.3 \\ 
    & 0.9999 & -19.8 $\pm$ 16.4 & -1.6 & 43.5 $\pm$ 33 & 6.8 & 6.6 $\pm$ 11.5 &  \colorbox{Goldenrod}{1.4} \\
    \hline
    \end{tabular}
    \label{tab:Different_gamma}
\end{table}

\begin{table}[htbp]
    \centering
      \caption{The best results of the SAC Agents with different discount factors $\gamma$ and entropy trade-off coefficients $\alpha$. \textit{Average} corresponds to the highest score and the lowest maximum force and number of relaxation steps of the averaged learning curves, marked with circles in Fig. \ref{fig:Different_gamma_and_alpha_SAC}. \textit{Best} represents the top-performing results among all seeds used for averaging. The highest scores and the lowest maximum force and number of relaxation steps are highlighted.}
    \begin{tabular}{ | c | c | c | c | c | c | c | c | }
     \hline
      \multicolumn{2}{|c|}{} & \multicolumn{2}{|c|}{Score, eV/A} & \multicolumn{2}{|c|}{Relaxation  steps} & \multicolumn{2}{|c|}{Max force, eV/A} \\
     \hline
       $\gamma$ & $\alpha$ & Average & Best & Average & Best & Average &  Best \\
    \hline
    \multirow{ 4}{*}{0.9} & 0.2 & -76.3 $\pm$ 37.6 & -17.2 & 78 $\pm$ 11 & 46.2 & 0.44 $\pm$ 0.28 & 0.01 \\
    
    & 0.1 & -48.3 $\pm$ 19.6 & -13.2 & 66 $\pm$ 16 & 34.4 & 0.25 $\pm$ 0.21 & 0.005 \\
    
    & 0.05 & -25.8 $\pm$ 5 & -6.3 & 52 $\pm$ 32 & 27.5 & 0.06 $\pm$ 0.07 & 0.004 \\
    
    & 0.01 & \colorbox{Goldenrod}{-7.5 $\pm$ 5.5} & \colorbox{Goldenrod}{-2.3} & \colorbox{Goldenrod}{23.3 $\pm$ 9.3} & \colorbox{Goldenrod}{9.1} & \colorbox{Goldenrod}{0.006 $\pm$ 0.001} & 0.004 \\
    
    \hline
    
    \multirow{ 4}{*}{0.99} & 0.2 & -34.4 $\pm$ 4.4 & -10.1 & 58.4 $\pm$ 13.8 & 25.5 & 0.09 $\pm$ 0.1 & 0.005 \\
    
    & 0.1 & -26.2 $\pm$ 11.3 & -5.9 & 57 $\pm$ 12 & 22.2 & 0.1 $\pm$ 0.08 & 0.004 \\
    
    & 0.05 & -79.8 $\pm$ 90.3 & -8.9 & 58 $\pm$ 30 & 22.7 & 0.7 $\pm$ 0.9 & \colorbox{Goldenrod}{0.003} \\
    
    & 0.01 & -14.3 $\pm$ 7.8 & -3 & 27.6 $\pm$ 7.9 & 10.0 & 0.012 $\pm$ 0.008 & 0.004 \\
    
    \hline
    
    \multirow{ 4}{*}{0.9999} & 0.2 & -154 $\pm$ 148.3 & -18.8 & 63 $\pm$ 27 & 37.6 & 1.2 $\pm$ 1.5 & 0.005 \\
    
    & 0.1 & -144 $\pm$ 181.5 & -6.8  & 60 $\pm$ 28 & 23 & 1.3 $\pm$ 1.8 & 0.005 \\
    
    & 0.05 & -64.4 $\pm$ 61.8 & -4.7 & 51 $\pm$ 28 & 16.8 & 0.09 $\pm$ 0.1 & \colorbox{Goldenrod}{0.003} \\
    
    & 0.01 & -16.3 $\pm$ 8.2 & -5 & 45 $\pm$ 19 & 15 & 0.01 $\pm$ 0.004 & \colorbox{Goldenrod}{0.003} \\
    
    \hline
    \end{tabular}
    \label{tab:Different_gamma_and_alpha_SAC}
\end{table}

\section{Structures and potentials}\label{Structures}

The model was trained on Al and Fe binary configurations and monoatomic Al and Fe configurations. For the calculation of forces, energies, and stress tensors was chosen Embedded-Atom Method (EAM).  For calculations with $N \leq 4$ we used structured form Materials Project while for higher $N$ we generated structures using USPEX code \cite{USPEX} and randomly chose one configuration for each $N$. All structures along with their Materials Project ID \cite{MatProj} (if such exist) and links to potentials are listed in Table \ref{Table_struct}. 

\begin{table}[h]
\centering
\caption{Structures used for model training. Here $N$ is the number of atoms in initial unit cell. During training structures with $N=1$ are doubled so $2\times1\times1$ supercell is relaxed.}
\begin{tabular}{ || l| c  | c | c | c | c || }
\hline
& Dataset & Structures &  Materials Project ID & N & EAM\\\hline
0  & \multirow{11}{*}{}  & alpha-U-type (Cmcm) Fe & mp-1271128 &2 &\multirow{7}{*}{\cite{Fe_EAM}}\\
1 & & b.c.c. Fe & mp-13 & 2 &\\
2 & & I4/mmm-distorted b.c.c Fe  & mp-1271068 & 4 & \\
3 & & Simple hexagonal Fe  & mp-1096950 & 1 & \\
4 & & h.c.p Fe  & mp-136 & 2 &\\ 
5 & Monoatomic & hypothetical I4/mmm Fe  & mp-1271198 & 4 & \\ 
6 & Al and Fe & f.c.c. Fe & mp-150 & 1 & \\  \cmidrule{3-6}
7 & & h.c.p. Al & mp-2647008 & 2 & \multirow{5}{*}{\cite{Al_EAM}}\\ 
8 & & hypothetical I4/mmm Al & mp-1239196 & 4 &\\
9 & & Al & mp-998860 & 1 &\\
10 & & double h.c.p Al  & mp-1183144 & 4 &\\
11 & & f.c.c. Al & mp-134 & 1 &\\ \hline
12 & & CsCl-type AlFe & mp-2658 &  2 & \cite{AlFe_EAM} \\ \hline
13 & & hexagonal AlFe & mp-985578 &  2 &  \\\hline
14 & & AlFe$_7$  & - &  8 & \cite{AlFe_EAM}\\\hline
15 & & Al$_8$Fe$_2$  & - &  10 & \cite{AlFe_EAM}\\\hline
16 & & Al$_2$Fe$_{18}$  & - &  20 & \cite{AlFe_EAM}\\\hline
\multirow{2}{*}{17} & \multirow{2}{*}{} Al-Fe compounds  & \multirow{2}{*}{-}  & \multirow{2}{*}{-} & \multirow{2}{*}{2-3}  &  \multirow{2}{*}{\cite{Fe_EAM}\cite{AlFe_EAM}\cite{Al_EAM}}\\
 & with $N =2-3$ &  & &  & \\\hline
 \multirow{2}{*}{18} & \multirow{2}{*}{} Al-Fe compounds  & \multirow{2}{*}{-}  & \multirow{2}{*}{-} & \multirow{2}{*}{4}  &  \multirow{2}{*}{\cite{Fe_EAM}\cite{AlFe_EAM}\cite{Al_EAM}}\\
 & with $N =4$ &  & &  & \\\hline
  \multirow{2}{*}{19} & \multirow{2}{*}{} Al-Fe compounds  & \multirow{2}{*}{-}  & \multirow{2}{*}{-} & \multirow{2}{*}{5}  &  \multirow{2}{*}{\cite{Fe_EAM}\cite{AlFe_EAM}\cite{Al_EAM}}\\
 & with $N =5$ &  & &  & \\\hline
\hline
\end{tabular}
\label{Table_struct}
\end{table}

\section{Hyperparameters}\label{Hyperparams}

For all models Replay size = $10^6$, learning rate of the Critic model optimizer $\lambda_c = 10^{-5}$, learning rate of the Actor model optimizer $\lambda_a = 10^{-5}$, batch size $= 100$, target noise for TD3 Agents $ = 0.05$, noise clip for TD3 Agents $ = 0.1$, policy delay $ = 2$, update every $= 1$, update after $= 100$,  $r_0 = $ 1.5 A. All models had 3 convolutional layers. $l^{(a)}_{max} = l^{(c)}_{max} = 3$ for all models except on Fig. \ref{fig:CGCNN_vs_E3NN}, where  $l^{(a)}_{max} = l^{(c)}_{max} = 2$. For Fig.  \ref{fig:CGCNN_vs_E3NN},  Fig. \ref{fig:Different_rewards} and Fig. \ref{fig:Different_rewards_action} start step was 500, for others it was 0. Noise setting for each model except on Fig. \ref{fig:CGCNN_vs_E3NN} and  Fig. \ref{fig:Different_rewards_action} is \textit{noise to state}, while for Fig. \ref{fig:CGCNN_vs_E3NN}, Fig. \ref{fig:Different_rewards_action} and  Fig. \ref{fig:Different_rewards_action} it is \textit{noise to action}. In all experiments $\beta = 0.5$, except Fig. \ref{fig:CGCNN_vs_E3NN} and  Fig. \ref{fig:Different_rewards}, where $\beta = 0.05$. 

\begin{table*}
    \caption{Hyperparameters, used in the given work. Here F, ST, LF, and H denote \textit{force}, \textit{step}, \textit{log force}, and \textit{hybrid} reward function settings correspondingly. In the case of \textit{hybrid} reward function weights, introduced in Eq. (\ref{R4_atom}), were always $w_1 = 1, w_2 = 0, w_3=0.5$. Noise stands for noise level $\lambda$ in the case of TD3 algorithm and for trade-off coefficient $\alpha$ in the case of SAC algorithms. The Dataset row lists the structure numbers from the Table \ref{Table_struct} used for the input dataset.}
	\centering 
    \begin{tabular}{ |c|c|c|c|c| } 
    \hline
    Parameter & Fig. \ref{fig:CGCNN_vs_E3NN} & Fig. \ref{fig:Different_rewards}, Fig. \ref{fig:Different_rewards_action}  & Fig. \ref{fig:Al_step}a) & Fig. \ref{fig:Al_step}b) \\
    \hline
    $\gamma$ & 0.99 & 0.9 (F, ST), 0.5 (LF)  & 0.9999 & 0.9999\\ 
    Noise & $[10^{-2},10^{-3}]$ & $[10^{-2}, 10^{-3}]$ & 0.2 & 0.2 \\
    Reward  & F & F, ST, LF & H, H + ST & F, F + ST\\
    $\epsilon_1$, eV/A  & $10^{-6}$ & $10^{-4}$ & $10^{-2}$  & $10^{-1}$ \\
    Dataset  & $\#$12 & $\#$12 & $\#$8  & $\#$0-11  \\
    \hline
    \hline
    & Fig. \ref{fig:Gener_from_AlandFe_to_AlFe}b-c) &  Fig. \ref{fig:Gener_from_AlandFe_to_AlFe}a) & Fig. \ref{fig:Al_andFe_supercell} & Fig. \ref{fig:Active} \\ \hline
    $\gamma$ & 0.9999 & 0.9999 & 0.9999  & 0.9999 \\ 
    Noise & 0.2 & 0.2 & 0.2 & 0.2 \\
    Reward & F & F, H + ST& H + ST & H \\
    $\epsilon_1$, eV/A  & $10^{-2}$ & $10^{-2}$, 0.2, 0.25 & $10^{-2}$, 0.2, 0.25 & - \\
    Dataset  & $\#$0-11 & $\#$12, $\#$8, $\#$14-16  & $\#$8, $\#$14-16 & $\#$17-19    \\
    \hline
    \hline
    & Fig. \ref{fig:Sens10} &  Fig. \ref{fig:Sev_str_greedy_vs_norm}a-c) & Fig. \ref{fig:Sev_str_greedy_vs_norm}d-f) & Fig. \ref{fig:Different_gamma_and_noise_SAC_TD3} \\ \hline
    $\gamma$ & 0.9999 & 0.9999 & 0.9999 & -   \\ 
    Noise & 0.2 & 0.2 & 0.2 & - \\
    Reward  & H & F & H, F& F \\
    $\epsilon_1$, eV/A  & 0.2  & $10^{-1}$  & $10^{-2}$ & $10^{-2}$  \\
    Dataset & $\#$18 & $\#$0-11  & $\#$8 & $\#$12   \\ 
    \hline
    \hline
    & Fig. \ref{fig:Different_gamma} & Fig. \ref{fig:Different_gamma_log_force} &  Fig. \ref{fig:Different_gamma_and_alpha_SAC} &    \\ \hline
    $\gamma$ & - & - & - &   \\ 
    Noise & - & [0.01, 0.001] & - &  \\
    Reward  & F & LF  & F &  \\
    $\epsilon_1$, eV/A  & $10^{-2}$ & $10^{-4}$  & $10^{-2}$ &   \\
    Dataset & $\#$12  & $\#$12  & $\#$12 &    \\ 

    \hline
    \end{tabular}
\end{table*}

For all TFN models $mul = 20$, number of  neighbors = 25, $em dim =10$, number of basis=10, radial layers=1, radial neurons=100.

\end{appendices}

\bibliographystyle{unsrt}  


\end{document}